\documentclass{aa}  

\usepackage{hyperref}
\usepackage[dvipsnames]{xcolor} 
\hypersetup{
    colorlinks=true,
    linkcolor=blue,
    filecolor=magenta,
    citecolor=teal,      
    urlcolor=cyan,
    }

\usepackage{float}
\usepackage{graphicx}
\usepackage{txfonts}
\usepackage{lipsum}
\usepackage{subcaption}         
\usepackage{lscape}             
\usepackage{placeins}           

\begin{document}

   \title{FASTAR - II. Semi-resolved evolutionary stellar population models}


   \author{Ignacio Martín-Navarro\inst{1,2}, Alexandre Vazdekis\inst{1,2}, Luis Peralta de Arriba\inst{3}, Isaac Alonso Asensio\inst{1,2}, Patricia Iglesias Navarro\inst{1,2}, Eirini Angeloudi\inst{1,2}, Francesco La Barbera\inst{4},  Miguel Cerviño\inst{5}, Katja Fahrion\inst{6}, Tereza Jerabkova\inst{7}, Michael A. Beasley \inst{1,2,8}, Jes\'us Falc\'on-Barroso\inst{1,2}, Marc Huertas-Company\inst{1,2,9,10}, Sebasti\'an F. S\'anchez\inst{11,1,2}, Prashin Jethwa\inst{5} }

   \institute{
Instituto de Astrof\'{\i}sica de Canarias,c/ V\'{\i}a L\'actea s/n, E38205 - La Laguna, Tenerife, Spain\\
\email{imartin@iac.es}
\and
Departamento de Astrof\'isica, Universidad de La Laguna, E-38205 La Laguna, Tenerife, Spain
\and 
Departamento de Inteligencia Artificial, Universidad Nacional de Educaci\'on a Distancia (UNED), Calle Juan del Rosal 16, E-28040 Madrid, Spain
\and
INAF-Osservatorio Astronomico di Capodimonte, sal. Moiariello 16, Napoli 80131, Italy
\and
Centro de Astrobiología (CSIC/INTA), 28692 ESAC Campus, Villanueva de la Cañada, Madrid, Spain
\and
Department of Astrophysics, University of Vienna, Türkenschanzstraße 17, 1180 Wien, Austria
\and
Centre for Astrophysics and Supercomputing, Swinburne University, John Street, Hawthorn, VIC 3122, Australia
\and 
Department of Theoretical Physics and Astrophysics, Faculty of Science, Masaryk University, Kotlářská 2, Brno 611 37, Czech Republic
\and
Observatoire de Paris, LERMA, PSL University, 61 avenue de l'Observatoire, F-75014 Paris, France
\and 
Universit\'e Paris-Cit\'e, 5 Rue Thomas Mann, 75014 Paris, France
\and
Instituto de Astronom\'ia, Universidad Nacional Aut\'onoma de M\'exico, A.P. 106, Ensenada 22800, BC, M\'exico
}

   \date{Received; accepted}
   \titlerunning{FASTAR II}
   \authorrunning{Martín-Navarro et al.}
 
  \abstract
   {
    Standard evolutionary synthesis models rely on the assumption of a fully sampled stellar initial mass function (IMF). Under this assumption, the age, chemical composition, and IMF uniquely define the predicted absorption spectra. However, with current instrumentation pushing observations towards higher spatial resolutions and lower surface brightnesses, the assumption of a fully sampled IMF does not always hold true. Here we present the semi-resolved version of the FASTAR models, a comprehensive set of evolutionary synthesis predictions able to reproduce the stochastic behavior of discretely-sampled IMFs. Semi-resolved FASTAR predictions share the same evolutionary principles, ingredients, and features of the integral (fully sampled IMF) version of the FASTAR models, expanding a range of ages from 20 Myr to 14 Gyr, metallicities between $-2.5 \le$ [M/H] $\le +0.3$, and several IMF functional forms. Detailed spectroscopic measurements can be carried out within the $3,540$--$7,400$ $\AA$ wavelength range, and low-resolution spectral energy distributions can also be synthesized over a wider 2,000-to-12,000\ $\AA$ coverage. Semi-resolved FASTAR models also depend on the number of stars contributing to the observed spectra, which determines the effective sampling of the different stellar evolutionary phases along the isochrones. This incomplete sampling implies that semi-resolved FASTAR models are inevitably stochastic. On top of the inherent stochasticity of the models, derived quantities such as equivalent widths, colors, or mass-to-light ratios might present strong deviations compared to standard fully sampled simple stellar population models. This stochasticity dilutes the boundary between model predictions and data, promoting new sampling-based inference approaches. FASTAR semi-resolved models allow for the effective exploration of the parameter space thanks to their optimized, JAX-based computation.
   }
   \keywords{galaxies: evolution -- galaxies: stellar content-- stars: evolution}

\maketitle
\nolinenumbers 

\section{Introduction}
Understanding the formation and evolution of galaxies requires an accurate interpretation of their integrated light. Beyond the Local Group, where stars can be individually resolved, the photometric and spectroscopic properties of galaxies are commonly interpreted in terms of physically meaningful quantities via evolutionary stellar population synthesis models. While alternative approaches are possible \citep[e.g.,][]{Spinrad71,Faber72,Pickles,Bica86,Bica88,Schmidt91,Pelat98,MN24b} the synthesis of integrated stellar population models anchored in the predictions from stellar evolution theory is now one of the most widely adopted tools in extragalactic astronomy \citep[e.g.,][]{Tinsley76,Worthey94,Leitherer99,bc03,TMB:03,Maraston05,Schiavon07,miles,Conroy12,Robotham25}. 

Simple stellar population (SSP) models are the fundamental output of evolutionary synthesis codes. An SSP model aims to represent the flux (luminosity) emitted by an (infinitely large) collection of stars with the same age and chemical composition. With the relative number of stars with different masses determined by the stellar initial mass function (IMF), the predicted spectral energy distribution of an SSP model is uniquely defined by the age of the population, its chemistry, and the underlying IMF. Conversely, the observed spectro-photometric properties of galaxies can be translated into these basic parameters or related quantities such as stellar masses or star formation rates \citep[see e.g.,][for an in-depth review of the synthesis and potential uses of evolutionary models]{Conroy13}. 

Over the years, the refinement of SSP models has reached a great level of maturity, and their success is undeniable. Yet, their range of applicability is limited to situations where the fundamental assumptions of evolutionary synthesis hold. While this is true in a large fraction of real-world applications, some interesting astrophysical cases push standard SSP models beyond their limits. In particular, when the number of stars per resolution element (i.e., the number of stars contributing to the observed spectral energy distribution) is not large enough, evolutionary population synthesis becomes intrinsically stochastic \citep[e.g.,][]{Miguel04, Miguel06, Miguel13}. Therefore, in this so-called semi-resolved regime, model predictions are no longer uniquely defined by the age, chemical composition, and IMF of the stellar population.

The stochastic nature of finitely samples stellar populations is well known \citep[e.g.,][]{Buzzoni93,Renzini98,Miguel02,Conroy16}. From a practical perspective, it has mostly been applied to young stellar populations in which the presence or absence of a few massive stars can significantly affect the integrated stellar spectrum of a population \citep[e.g.,][]{dS12,Eldridge12,Krumholz15,Orozco-Duarte,Stanway23}. In addition, the stochasticity in the integrated light of star clusters has also been the subject of dedicated studies \citep{Miguel00,Bruzual02,Fouesneau10,Fouesneau12,Beerman12,Branco24}. Moreover, the underlying physical process enabling the use of surface brightness fluctuations as distance \citep{Tonry88,Tonry01,Blakeslee09} and stellar population content \citep{Miguel08,Pablo21} indicators is also the inherently stochastic nature of semi-resolved populations.

Semi-resolved stellar populations are therefore a natural expectation. Up until recently, however, the study of them has been focused on the specific scientific cases described above, although current technological advances are pushing the limits out to where stochasticity cannot be neglected. For example, high spatial resolution studies of relatively nearby galaxies using integral field spectroscopy show fluctuations in the recovered stellar population maps larger than expected from the signal-to-noise of the data \citep[e.g.,][]{Pinna19b,Bittner20,Justus2020,MN21,LVM,Sebastian25S}.

More critically, the increasing sensitivity of optical detectors is pushing the limits of low-surface-brightness studies \citep[e.g.,][]{Abraham14,vdk15,Trujillo16,Mihos17,Trujillo21,MD25}. With forthcoming ground- and space-based facilities specifically devoted to obtaining unprecedentedly deep photometric data, such as the Dark Energy Spectroscopic Instrument \citep{desi}, Euclid \citep{Euclid0,Euclid}, or the Vera C. Rubin Observatory \citep{lsst,Martin22}, establishing robust methods to model and interpret semi-resolved observations will be crucial for fully exploiting the scientific potential of the low-surface-brightness Universe. Yet the most important challenge for standard stellar population models lies ahead. The milliarcsecond physical-scale resolution of the next generation of 40-meter-class telescopes, such as the Extremely Large Telescope, will yield semi-resolved observations of every nearby galaxy. In this context, detailed observational astronomy in the coming decades will be, by necessity, semi-resolved.

In the first paper of this series, we presented FASTAR, a differentiable evolutionary stellar population synthesis code. Here we make use of the fast and optimized  tools of FASTAR to describe the synthesis of semi-resolved predictions. These predictions are based on the exact same principles as standard SSP models, retaining their strength and applicability, but are made to match semi-resolved observations with a variable number of stars. The layout of this paper is as follows. In Sect.~\ref{sec:ingredients} we detail the basic FASTAR ingredients, in Sect.~\ref{sec:synthesis} we present the synthesis of the semi-resolved predictions, and in Sect.~\ref{sec:results} we outline the overall behavior of the models. In Sect.~\ref{sec:implications} we expand on some immediate applications of the FASTAR predictions, while in Sect.~\ref{sec:deal} we discuss some of the observational practicalities. Finally, in Sect.~\ref{sec:summary} we summarize the main features of the semi-resolved FASTAR models and outline upcoming developments.

\section{Model ingredients} \label{sec:ingredients}

FASTAR semi-resolved predictions are based on the same model ingredients as the standard fully sampled SSP models described in the first paper of this series. Semi-resolved predictions are also computed using the JAX Python library \citep{jax2018github,jax}, with native numerical autodifferentiation and optimized computation with both a CPU and GPU. The five main FASTAR ingredients are described in the following.

\subsection{Initial mass functions}
FASTAR comes with six predefined functional forms for the IMF. Models with both \citet{mw} and \citet{Chabrier} IMF shapes can be synthesized representing the Milky Way standard. In addition, FASTAR also incorporates a single and a broken power-law (three segments) IMF parameterization with variable slopes. Furthermore, the so-called bimodal IMF functional form described in \citet{vazdekis96} is also included as well as a tapered power-law, as defined in \citet{Guido05}. All the fixed and variable parameters of these IMF functional forms are easily accessible and modified according to the user's needs. More details on each IMF description are given in Section 2.1 of the first FASTAR paper. In addition to these predefined IMF parametrizations, FASTAR also offers the possibility of generating model predictions for any user-defined IMF functional form.

\subsection{Isochrones}
Stellar evolutionary theory is coded into the FASTAR models through the BaSTI-IAC isochrones \citep{Hidalgo18,Pietrinferni21,Pietrinferni24}. In particular, we used the latest version of their solar-scaled isochrones, including improved prescriptions for atomic diffusion and overshooting. This set of isochrones covers a range in age from 20 Myr to 14 Gyr and -2.5 to +0.3 in total metallicity [M/H], which in practice sets the applicability range of the models. 

Currently, FASTAR model predictions do not include variable elemental abundances, even if individual stars have non-solar abundances (see details below in the stellar libraries). As for the case of the standard FASTAR SSP predictions, semi-resolved FASTAR models can be evaluated at any specific age, metallicity, and IMF thanks to the consistent sampling of the BaSTI isochrones.

\subsection{Bolometric corrections}
Rather than using the bolometric luminosities of the BaSTI-IAC isochrones, we followed the synthesis approach of the MILES models \citep{miles,Vazdekis15}. In this approach, theoretical stellar luminosities are translated into V-band absolute magnitudes using the bolometric corrections of \citet{Worthey11}. FASTAR predictions are anchored to the Sun, assuming a bolometric magnitude of 4.70 and a V-band bolometric correction of BC$_\sun=-0.12$. This results in flux predictions scaled to a distance of 10 parsecs.

\subsection{Stellar interpolator}
The core component of FASTAR is its stellar interpolator which provides fast, accurate, and differentiable stellar spectra. In a nutshell, FASTAR interpolation follows a similar approach as in \citet{Alsing20}. Instead of predicting the entire spectrum of a star given its atmospheric parameters ($T_{\mathrm{eff}}$, $\log g$, [Fe/H]), we first reduce the dimensionality of the spectra using a simple principal component analysis (PCA) decomposition. This initial step reversibly compresses the entire spectrum (usually evaluated at a few thousand wavelengths) into, in our case, a 16-element vector, containing the PCA projection coefficients.

Then, a simple fully connected neural network with four hidden layers (64, 128, 128, 64) is trained using a mean squared error loss function to predict these 16 coefficients given the three atmospheric parameters. As demonstrated in the first FASTAR paper, this interpolation scheme performs at the same level as the stellar interpolator behind the MILES models. However, it is orders of magnitude faster in the synthesis of new stellar spectra.

\subsection{Stellar libraries}
Using the ingredient above, FASTAR can produce two sets of predictions. On the one hand, over the $3,540$--$7,400$ $\AA$ wavelength range, FASTAR provides detailed spectroscopic predictions at a 2.51 \AA \ full width at half maximum resolution \citep[same as MILES][]{Jesus11}. These predictions are based on the combination of two stellar libraries: the Medium-resolution Isaac Newton Telescope library of empirical spectra \citep[MILES,][]{Pat06} and the BOSZ synthetic stellar spectral library \citep{Meszaros24}. Details on the combination and training of the PCA-based neural network regressor are given in the first FASTAR paper, where we also demonstrate how the combination of both stellar libraries yields accurate stellar predictions across the entire $T_{\mathrm{eff}}$--$\log g$--[Fe/H] parameter space. Powered by the thoroughly tested MILES stellar templates, this set of FASTAR predictions is meant to be used for detailed spectroscopic analyses.

For many photometric applications, however, a broader wavelength range is required. Therefore, FASTAR has an additional set of predictions that is entirely based on the BOSZ theoretical stellar templates, expanding from 2,000 to 12,000 \AA. These alternative FASTAR models can be convolved with any set of photometric filters to reproduce generic spectral energy distributions but are intentionally undersampled in the spectral direction ($\Delta \lambda = 4$ \AA) to discourage their use for spectroscopic purposes.

\begin{figure*}
    \centering
    \includegraphics[width=16cm]{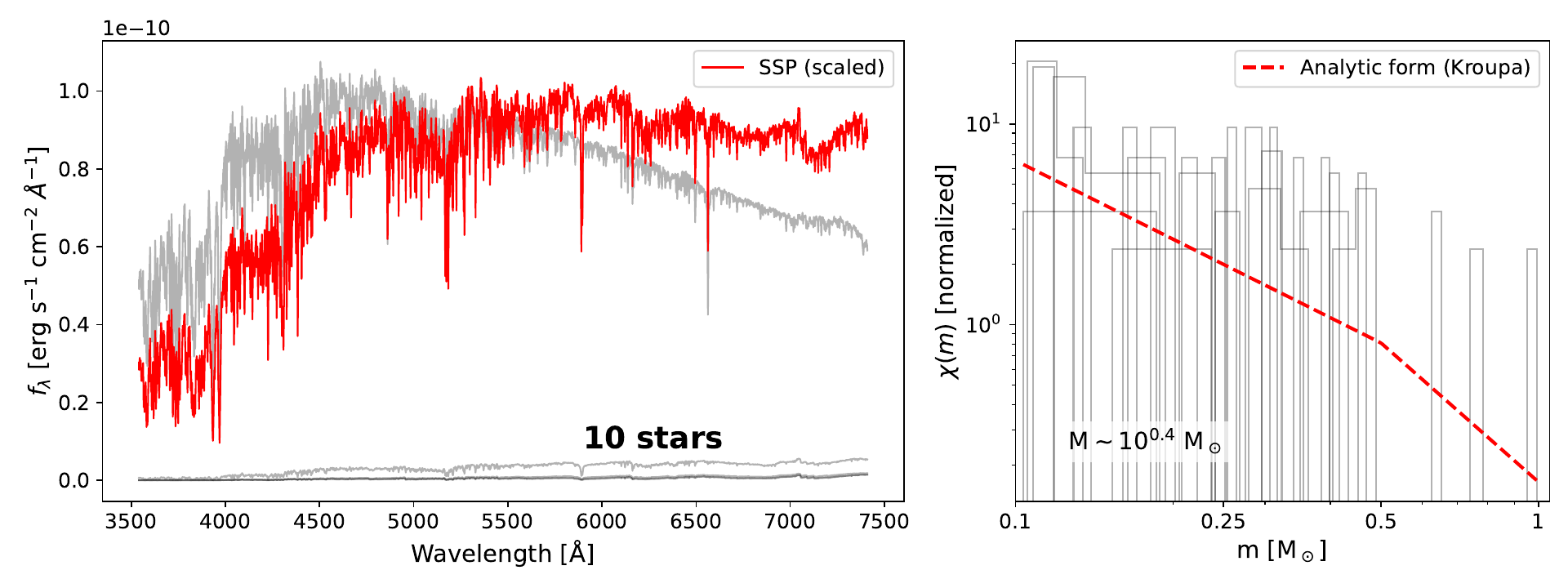}
    \includegraphics[width=16cm]{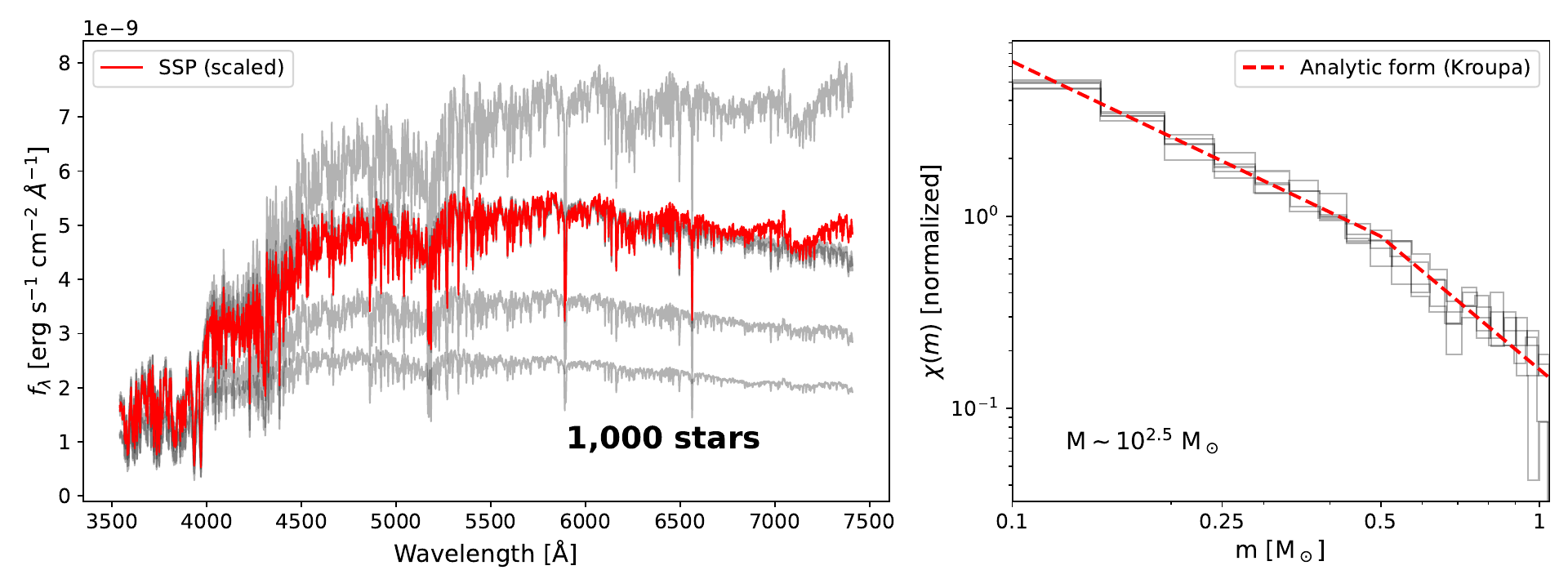}
    \includegraphics[width=16cm]{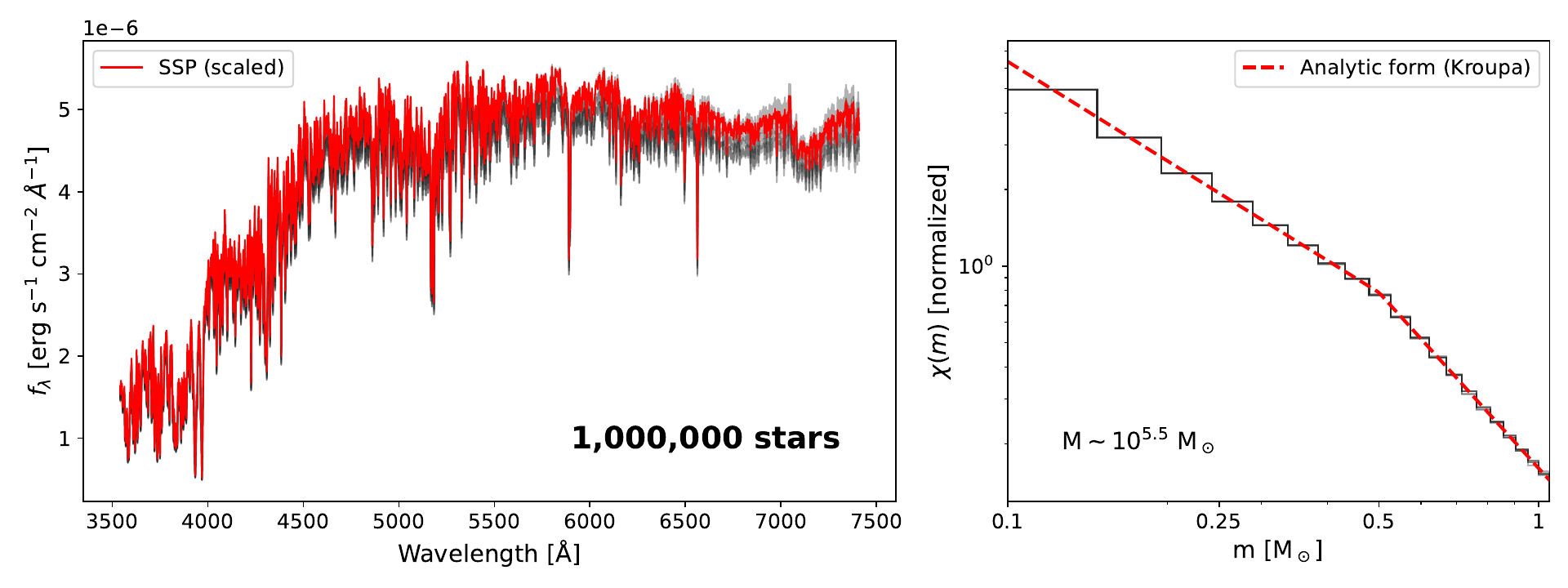}
    \caption{Stochastic nature of FASTAR semi-resolved predictions. In the left panel, each row shows five random semi-resolved SSP models (in gray) resulting from different stochastic samplings of the IMF. The corresponding IMF realizations are shown in the right panels (also in  gray), along with the assumed IMF (red) and the mean stellar mass of the population. For reference, the standard fully sampled SSP prediction is included in red in the left panels (scaled to arbitrary units for comparison). All realizations correspond to a 10 Gyr, solar-metallicity population. From top to bottom, the number of stars increases from 10 to 1,000,000.}
    \label{fig:semi_ssp}
 \end{figure*}

\section{Semi-resolved versus integral synthesis} \label{sec:synthesis}

In general, standard evolutionary stellar population models define the flux emitted by an SSP with a given age and metallicity as

\begin{equation} \label{eq:1}
    F_\lambda \bigl(\text{age}, [\mathrm{M}/\mathrm{H}] \bigr) 
    = \int_{m_{\mathrm{low}}}^{m_{\mathrm{high}}(age)} 
      S_\lambda^{M_V}\!\left(m \mid \mathrm{age},[\mathrm{M}/\mathrm{H}]\right)
      \,\chi(m) \, dm
\end{equation}

\noindent
where $S_\lambda^{M_V}\!\left(m \mid \mathrm{age},[\mathrm{M}/\mathrm{H}]\right)$ are the interpolated stellar spectra (scaled to their expected V-band absolute magnitude) and $\chi(m)$ the relative number of stars with different stellar masses, i.e., the IMF. Underneath Eq.~\ref{eq:1}, stellar evolution theory determines the integration limits (in particular the most massive star contributing to the observed flux), the mapping between stellar mass, age, metallicity and $T_{\mathrm{eff}}$--$\log g$ (i.e., the isochrones) and the luminosity or scaling factor of each $S_\lambda^{M_V}$.

Equation~\ref{eq:1}, however, only holds valid in the limit where a large enough number of stars are observed within a given resolution element\footnote{In the optical range, $N_\mathrm{stars}\sim10^5$ typically separates fully sampled from semi-resolved regimes \citep[see e.g.,][and sections below]{Miguel04}.}. To generate evolutionary semi-resolved predictions, $f_\lambda$, in FASTAR we modeled the flux of a finite ensemble of stars using a simple summation:

\begin{equation} \label{eq:discrete}
    f_\lambda\bigl(\text{age}, [\mathrm{M}/\mathrm{H}], N_\mathrm{stars}\bigr)
    = \sum_{i=1}^{N_\mathrm{stars}} S_\lambda^{M_V}\!\left(m \mid \mathrm{age},[\mathrm{M}/\mathrm{H}]\right).
\end{equation}

Similarly to standard SSP models, stellar evolution theory determines the relation between stellar mass and atmospheric parameters for a given age and metallicity and thus the stellar spectra ($S_\lambda^{M_V}$) on the right-hand side of Eq.~\ref{eq:discrete}. These stellar spectra are again scaled to match the V-band absolute magnitude predicted by the isochrones (modulo the bolometric correction). The main difference between the integral synthesis (Eq.~\ref{eq:1}) and the semi-resolved version (Eq.~\ref{eq:discrete}) is that in the latter, the mass of each of the $N_\mathrm{stars}$ is randomly sampled following the assumed IMF.

To illustrate the similarities and differences between integrated and semi-resolved FASTAR predictions, Fig.~\ref{fig:semi_ssp} shows the effect of increasing the number of stars, from 10 to 10,000,000, in an integrated stellar population. In FASTAR, the semi-resolved regime inherits exactly the same stellar population parameters as the standard integrated SSP models, namely age, total metallicity, and IMF, as these quantities define the evolutionary side of the model. On top of these, semi-resolved predictions in FASTAR are characterized by an additional parameter: the number of stars contributing to the integrated flux.

Although the age, metallicity, IMF, and number of stars fully specify the physical model, semi-resolved FASTAR predictions remain stochastic by construction. In other words, a fixed tuple of model parameters (age, [M/H], IMF, $N_\mathrm{stars}$) does not generate a unique spectrum since the IMF, and thus stellar evolutionary phases, are stochastically sampled. This is the fundamental distinction with respect to standard fully sampled evolutionary synthesis models: while the underlying physics and ingredients are identical, semi-resolved models produce a distribution of possible spectroscopic predictions rather than a single deterministic one.

An important consideration is worth highlighting here. For a given age and metallicity, stellar mass maps stellar evolution in a mono-parametric way, thus defining the properties of the individual stellar spectra that go into the synthesis of an SSP and their luminosities. Therefore, although FASTAR models are calculated through a random mass sampling of the IMF, the synthesized spectra effectively depend on the different stellar evolutionary phases probed by the isochrone. In simpler terms, the presence of a luminous red giant will have a much more noticeable impact on a semi-resolved SSP model that the absence of a low-mass M dwarf. Thus, the variety of spectral properties in the semi-resolved regime results from a complex and  nonlinear interaction between the number of stars, their stellar masses, and their evolutionary phase, all of this modulated by the assumed IMF. The upcoming sections will expand on this connection between number of stars, IMF sampling and stellar evolutionary phases.

The intrinsic stochasticity of these models can be quantified in different ways, for instance, by providing the mean spectrum (i.e. the SSP) and its variance for a grid of ages and metallicities \citep{Vazdekis20}. However, FASTAR offers a more natural approach to modeling the light from semi-resolved populations by explicitly sampling the IMF as many times as needed ($N_\mathrm{stars}$) to reproduce the observed data. Note that semi-resolved FASTAR models converge to the fully sampled IMF predictions as $N_\mathrm{stars} \to \infty$, generalizing the use of evolutionary stellar population models.

Two immediate things are evident from Fig.~\ref{fig:semi_ssp}. First, the most obvious one, the scatter decreases as the number of stars increases. It is worth noting also that, as expected, the convergence towards the mean SSP value (red spectra) happens faster towards bluer wavelengths at these old ages \citep[see e.g.,][]{Miguel13}. Second, on average, the total luminosity increases as well with the number of stars. Contrary to standard fully sampled predictions that are usually scaled to one solar mass,  semi-resolved FASTAR models predict the actual flux, scaled at 10 pc, expected from a population with the given number of stars and IMF sampling. 

\section{Model predictions} \label{sec:results}
Despite the differences between fully sampled and semi-resolved evolutionary model synthesis, the predicted quantities are the same. In particular, FASTAR semi-resolved models can be used to generate spectra as demonstrated in Fig.~\ref{fig:semi_ssp}, which can then be analyzed through line-strength indices and colors as traditional models. In addition, mass-to-light ratios can also be retrieved for each realization for any filter within the FASTAR wavelength coverage.

In the subsections below we describe how these derived quantities change as a function of the number of stars. The following figures also represent the distribution of expected values, indicating the 1/99 and 10/90 percentiles. Contrary to the standard way of delivering evolutionary model predictions, FASTAR allows the user to synthesize any arbitrary number of models. Thus, these percentiles boundaries are not designed to formally describe the behavior of the line strengths, colors etc. but to exemplify the range of possible values. Note also that the variation in these quantities should not be interpreted as a model uncertainty but as a consequence of the intrinsic stochasticity of semi-resolved evolutionary stellar population models.

\subsection{Line-strengths indices}

\begin{figure}
    \centering
    \includegraphics[width=7.cm]{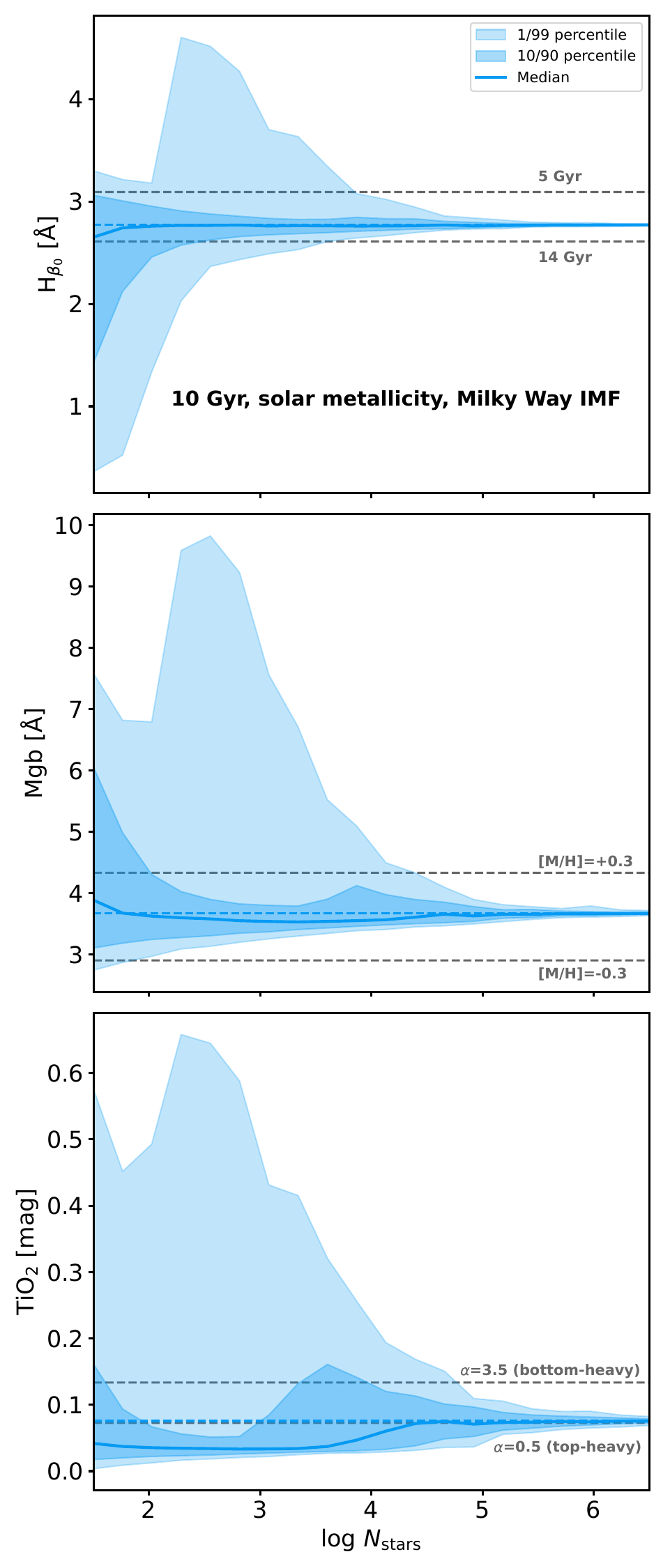}
    \caption{Behavior of line-strength indices. From top to bottom, the distribution of FASTAR semi-resolved predictions as a function of the number of stars in the population is shown for the H$_{\beta_O}$, Mgb, and TiO$_2$ features, respectively. These distributions correspond to 1,000,000 stochastic IMF samplings of the same 10 Gyr old, solar metallicity, and Milky Way-like IMF population. The median of the distribution is shown with a solid blue line and the shaded regions correspond to the 10/90 and 1/99 percentiles. The blue dashed line indicates the fully sampled SSP FASTAR prediction. For comparison, dashed horizontal lines mark the line-strength value expected in the fully sampled regime for various ages, metallicities and IMFs. Line-strength values have been calculated at the nominal 2.51 \AA \ resolution of the FASTAR models.}
    \label{fig:indices}
 \end{figure}

Figure~\ref{fig:indices} shows the behavior of the H$_{\beta_O}$ \citep[][top panel]{Cervantes}, the Mgb \citep[][middle panel]{Worthey94b}, and the TiO$_2$ \citep[][bottom panel]{Faber85,trager} line-strength indices as a function of the number of stars for a 10 Gyr old population with solar metallicity, and Milky Way-like IMF. These three indices are mostly sensitive to age, metallicity and IMF slope, respectively, and exemplify how line-strength predictions can vary in the semi-resolved regime even when the underlying population has the same physical properties. Shaded blue areas indicate the 1/99 and 10/90 distribution percentiles over 1,000,000 realizations, while the solid blue line shows the median value. We increase the number of realizations when decreasing the number of stars to capture the enhanced variability of the model predictions. For comparison, each panel also includes dashed horizontal lines corresponding to the line-strength values expected when changing the stellar population properties for a fully sampled SSP model. 

When the number of stars falls below $N_\mathrm{stars}\sim10^5$ per resolution element, the stochastic fluctuations of the models become the dominant factor, inducing changes in the observed line-strength indices much larger than those expected from variations in age, metallicity, or the IMF (dashed horizontal lines). The asymmetric distribution around the median value results from the sampling of specific stellar masses and thus evolutionary phases (e.g. low-mass main sequence, turn-off, or giant star) when changing the number of stars in the population. This is clearly exemplified by the TiO$_2$ behavior. This index is more prominent in the atmospheres of cool giant stars \citep[e.g.,][]{Spiniello2013}, and therefore the measured line strengths are biased towards low TiO$_2$ values if the number of stars is low. Only when the contribution of the more massive and less frequent giant stars rises through a larger number of stars, the distribution of TiO$_2$ values converges towards the fully sampled SSP expectation. Interestingly, neither H$_{\beta_O}$ nor Mgb exhibit clear biases as the number of stars decreases. In general, if the number of stars is low, line-strength measurements cannot be directly translated to physical parameters. For completeness, appendix~\ref{app:young} includes a similar assessment but for a young (0.25 Gyr) population.

\begin{figure}
    \centering
    \includegraphics[width=7.3cm]{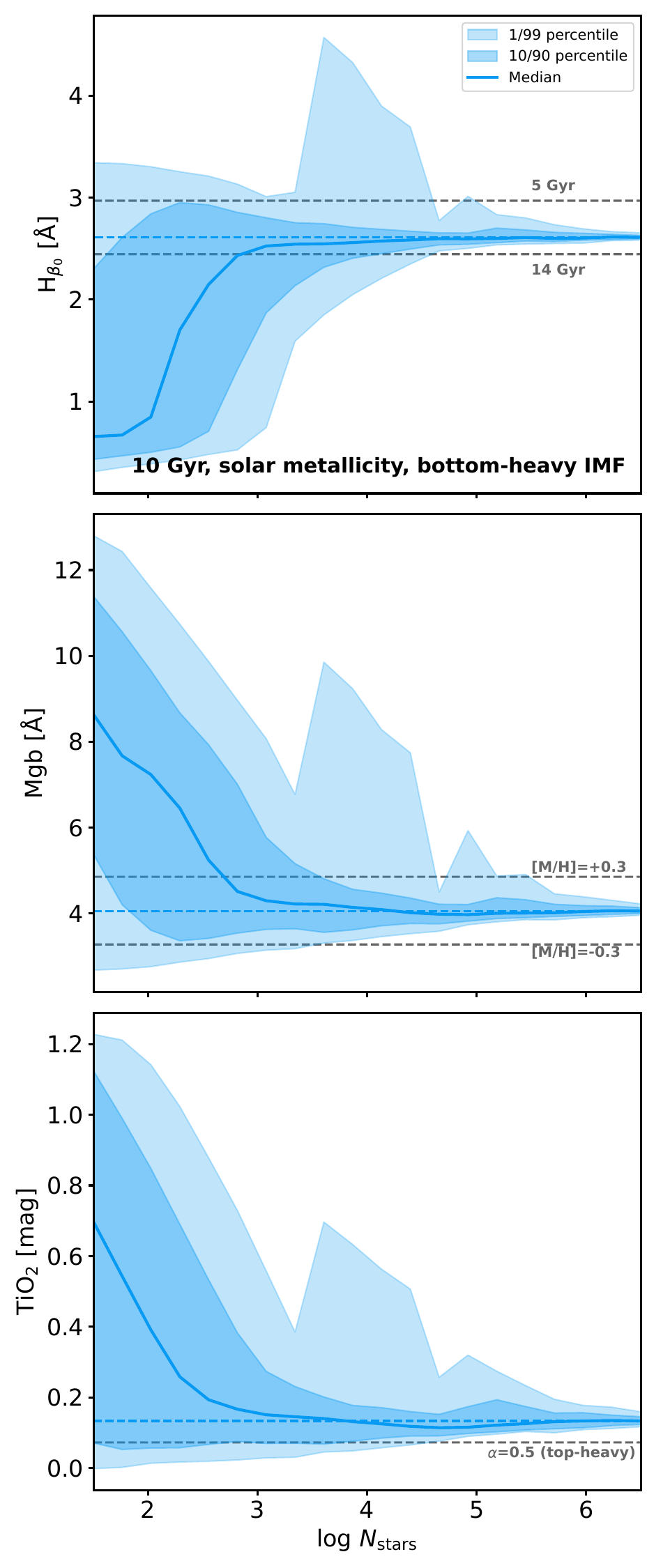}
    \caption{Same as Fig.~\ref{fig:indices} but line-strength predictions have been calculated assuming a single power law, bottom-heavy IMF ($\alpha=3.5$).}
    \label{fig:indices_bh}
 \end{figure}

It is evident from Fig.~\ref{fig:indices} that, in the semi-resolved regime, the recovered spectral properties are a balance between the intrinsic properties of the stellar population and the IMF and stellar evolution sampling, which depends on the number of stars per resolution element. Therefore, in this regime, the choice of the IMF has a strong effect on the model predictions. Figure~\ref{fig:indices_bh} is equivalent to Fig.~\ref{fig:indices} but in this case, the FASTAR semi-resolved predictions have been calculated assuming a single power law, bottom-heavy IMF (unimodal, $\alpha=3.5$). While there are obvious similarities between both figures, it is also clear that the IMF has an important role in modulating the behavior of the three indices. In particular, Fig.~\ref{fig:indices_bh} does show clear biases in the measured line-strength values when decreasing the number of stars, with weaker H$_{\beta_O}$ values and stronger Mgb and TiO$_2$ features as the number of stars decreases. 

\begin{figure}
    \centering
    \includegraphics[width=8cm]{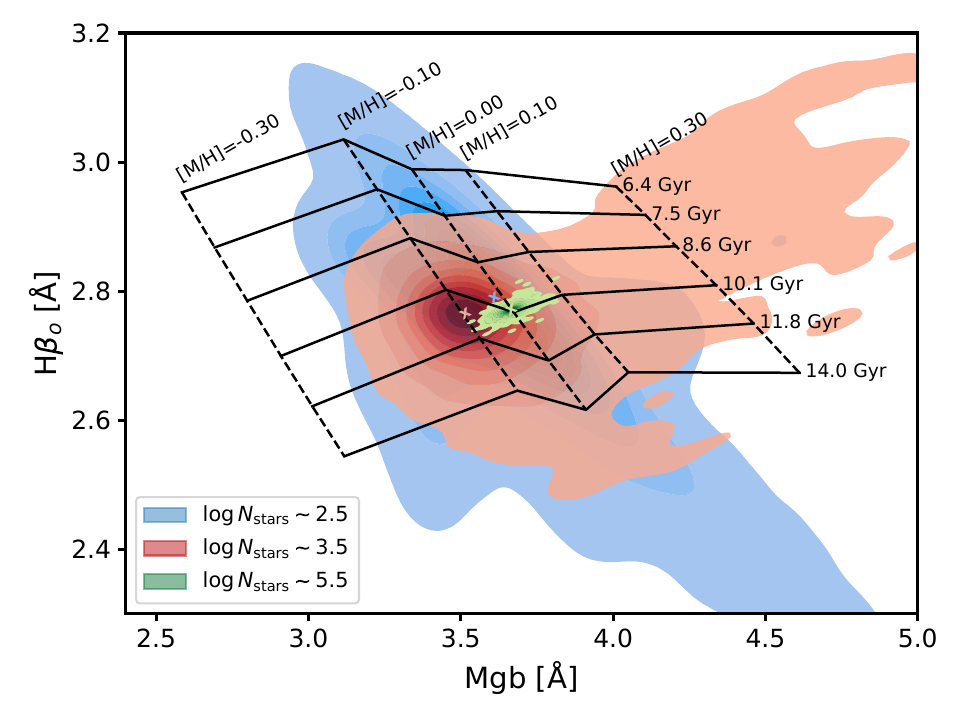}
    \caption{Semi-resolved index--index diagrams. Black lines correspond to the fully sampled H$_{\beta_O}$--Mgb FASTAR predictions for different ages and metallicities, as noted by the labels. On top, green, red, and blue contours show FASTAR index--index distributions for $N_\mathrm{stars}=10^{5.5}$, $10^{3.5}$, and $10^{2.5}$, respectively, for a population with solar metallicity and an age of 10 Gyr. Colored marks indicate the median of the distributions.}
    \label{fig:index_grid}
 \end{figure}

The intrinsic spread in semi-resolved spectral predictions, and consequently in the derived line-strength indices, places clear limits on how accurately stellar population parameters can be constrained from individual line-strength measurements\footnote{Note, however, that the shaded areas in Figs.~\ref{fig:indices} and \ref{fig:indices_bh} represent the expected range of model values and should not be interpreted as uncertainties in the recovered stellar population parameters. The latter will depend on the specific fitting approach and on the properties of the data.}. This effect is further illustrated in the H$_{\beta_O}$--Mgb grid shown in Fig.~\ref{fig:index_grid}. Such index--index diagrams are commonly used in detailed spectroscopic analyses \citep[e.g.,][]{Worthey92,Thomas05,Kuntschner10,LB19} since, in the fully sampled regime, a pair of line-strength measurements can (in principle) fully determine the properties of the underlying stellar population. These unique predictions for standard fully sampled SSP models are shown as a black grid in Fig.~\ref{fig:index_grid}.

In the semi-resolved regime, however, the prospects are different. The green, red, and blue contours in Fig.~\ref{fig:index_grid} show the projected distributions of line-strength measurements (from Fig.~\ref{fig:indices}) for $\log N_\mathrm{stars} = {5.5}$, ${3.5}$, and ${2.5}$, respectively. Even for a relatively large number of stars (green contours) the stochastic nature of finite, semi-resolved populations can introduce a scatter up to $\sim$0.1 dex in metallicity and $\sim$1 Gyr in age, despite all measurements corresponding to the same 10 Gyr, solar-metallicity model. For smaller numbers of stars per resolution element (red and blue contours), the scatter becomes much larger, with line-strength values extending well beyond the boundaries of the original model grid. In addition, small systematic biases emerge as the number of stars decreases.

\subsection{Colors}

\begin{figure}
    \centering
    \includegraphics[width=8cm]{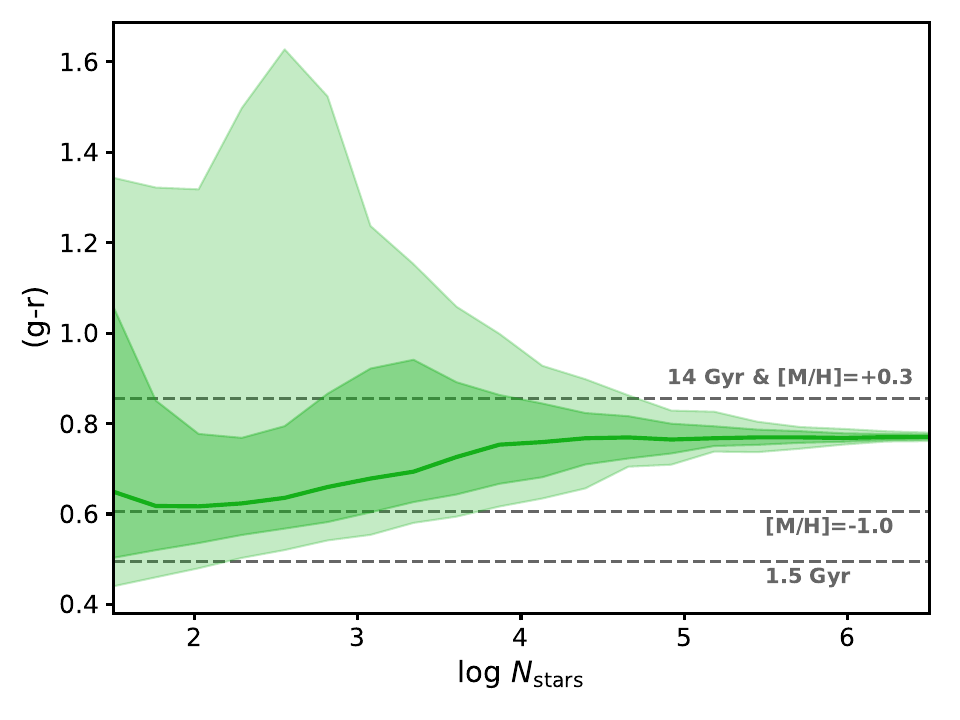}
    \caption{Color dependence on the number of stars. Green shaded areas indicate the 10/90 and 1/99 percentiles of the distribution of measured $(g-r)$ colors as a function of $N_\mathrm{stars}$ for semi-resolved populations of 10 Gyr, solar metallicity and a Milky Way-like IMF. The green solid line corresponds to the median trend. For reference, horizontal dashed lines mark the FASTAR color prediction for a fully sampled IMF with different ages and metallicities.}
    \label{fig:color}
 \end{figure}

Colors of semi-resolved populations can also be computed directly with FASTAR by convolving the model spectra with any desired set of photometric filters. As noted above, FASTAR models can be generated with full spectroscopic detail over the 3,540--7,400~\AA \ wavelength range, and for photometric applications over a broader 2,000–12,000~Å coverage. To illustrate the impact of stochastic IMF sampling in semi-resolved populations, Fig.~\ref{fig:color} shows the predicted $(g-r)$ color, as in Fig.~\ref{fig:indices}, calculated for a 10~Gyr, solar-metallicity population with a Milky Way-like IMF, using the narrower FASTAR wavelength range.

 The impact of stochasticity on the integrated colors of semi-resolved populations is larger than that observed in the line-strength indices. Even when the population is composed of $\sim 10^5$ stars, the 16/84 percentile variation ($\sim\pm1\sigma$) in the measured $(g-r)$ color is $\sim 0.09$ mag. When $N_\mathrm{stars}$ is even lower, the range of possible colors increases, becoming compatible, under the assumption of a fully sampled IMF, with a range of ages and metallicities, as indicated by the horizontal dashed lines in Fig.~\ref{fig:color}. On top of this large color scatter, a bias towards blue $(g-r)$ colors is also evident as the number of stars in the population decreases \citep[e.g.,][]{Miguel03}, reaching a minimum around $\log N_\mathrm{stars} \sim 2$.

\begin{figure}
    \centering
    \includegraphics[width=8cm]{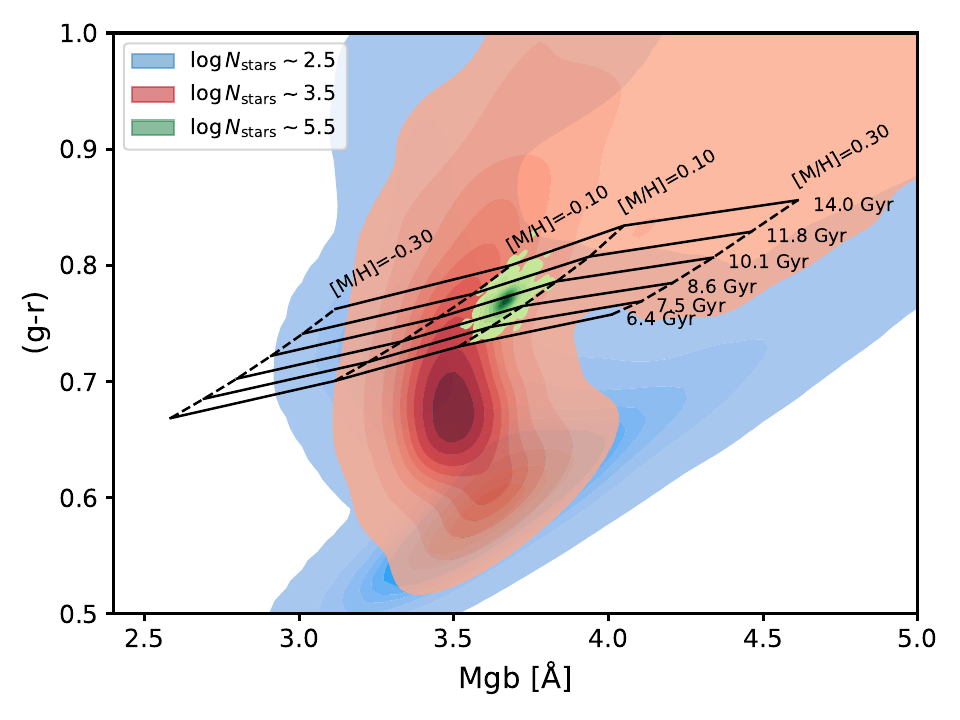}
    \caption{Color grid. Black lines indicate the FASTAR fully sampled predictions for the same age and metallicity ranges as in Fig.~\ref{fig:index_grid}. Green, red, and blue contours show the distributions of $(g-r)$ and Mgb values for a 10 Gyr and solar metallicity population with a variable number of stars ($N_\mathrm{stars}=10^{5.5}$, $10^{3.5}$, and $10^{2.5}$, respectively).}
    \label{fig:color_index}
 \end{figure}

 A large scatter in the color predictions for a given stellar population casts further doubts on the validity of standard analysis approaches when dealing with semi-resolved stellar systems. Figure~\ref{fig:color_index} shows, in black, the fully sampled FASTAR predictions for the $(g-r)$ vs Mgb plane and in colored contours (green, red, and blue) semi-resolved realizations with $\log N_\mathrm{stars}=5.5$, ${3.5}$, and ${2.5}$, respectively. The range of ages and metallicities of the grid is the same as in Fig.~\ref{fig:index_grid}.

 The increased scatter combined with the lesser sensitivity to changes in the stellar population properties of colors vs line-strengths is evident from Fig.~\ref{fig:color_index}. With populations as well sampled as $N_\mathrm{stars}\sim10^5$, the range of possible ages covers from $\sim6$ to $\sim14$ Gyr. When the number of stars drops below that threshold, the $(g-r)$--Mgb loses almost all predictive power, with a large spread of possible solutions and a clear bias towards blue colors.

\subsection{Mass-to-light ratios}

In conjunction with the properties of the stellar population (age, metallicity and IMF in the case of the FASTAR models), the number of stars is the fundamental parameter determining the resulting spectra of semi-resolved populations. In the fully sampled regime, the number of stars and the observed mass-to-light ratio are rigidly tied through the assumed IMF. Therefore, when the IMF is fully sampled, the observed luminosity of a stellar population can be directly translated into a stellar mass, or equivalently into a number of stars, through the expected mass-to-light ratio.

\begin{figure}
    \centering
    \includegraphics[width=8cm]{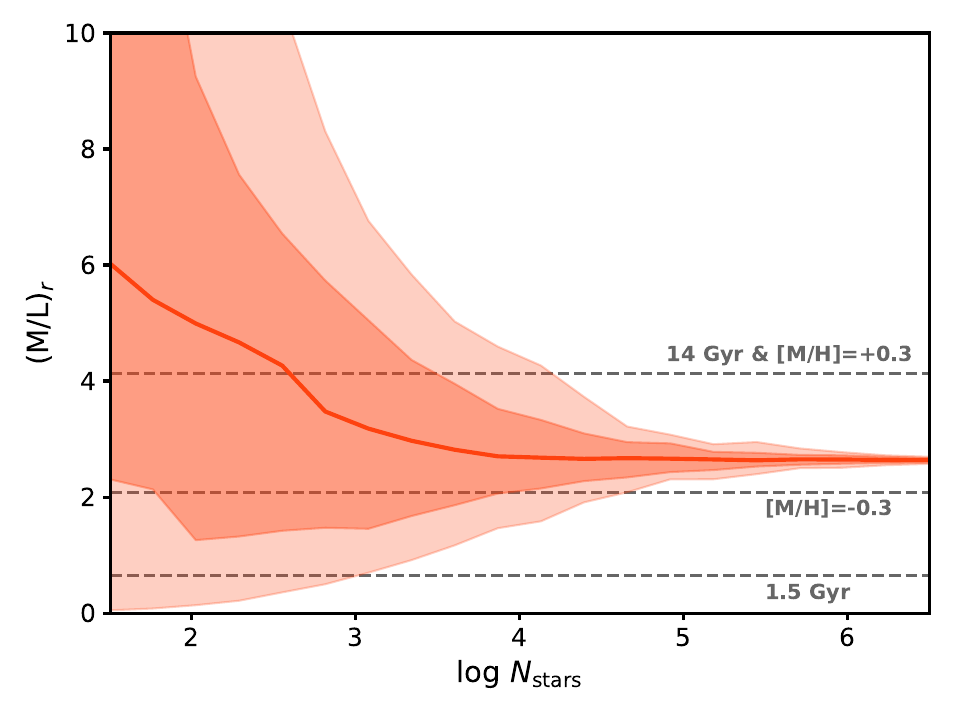}
    \caption{Mass-to-light ratio. The measured $r$-band mass-to-light ratio for a 10 Gyr and solar metallicity population is shown as a function of the number of stars in the semi-resolved FASTAR models. Shaded areas correspond to the 10/90 and 1/99 percentiles of the distribution of retrieved values after the stochastic sampling of the IMF. The solid red line indicates the median value for each $N_\mathrm{stars}$ bin.}
    \label{fig:ml}
 \end{figure}

However, in semi-resolved populations, the mapping between mass, luminosity, and number of stars breaks apart because of the stochastic sampling of the IMF. Illustrating how the mass-to-light ratio of a given population varies in the semi-resolved regime, Fig.~\ref{fig:ml} shows the predicted $r$-band mass-to-light ratio as a function of $N_\mathrm{stars}$, assuming again a 10 Gyr, solar metallicity and Milky Way-like IMF. As before, shaded areas denote the 10/90 and 1/99 percentiles of the distribution and the solid line indicates the median relation.

As in the previous figures, the intrinsic variability in the IMF sampling results in a wide range of actual mass-to-light ratios. While there is a clear bias towards high values as $N_\mathrm{stars}$ decreases, and therefore low-mass stars become more prominent with respect to higher and brighter stellar masses, the scatter can expand more than an order of magnitude, which is consistent with the expectations of both a very old and a very young population (horizontal dashed lines in Fig.~\ref{fig:ml}). 

Figure~\ref{fig:ml} showcases the difficulty of measuring the stellar mass of a semi-resolved stellar population based on its stellar population properties. Alternatively, mass-to-light ratios can also be estimated on the basis of scaling relations \citep[e.g.,][]{Bell01,Bell03,Zibetti09,McGaugh14}. In practice, this is done by plugging in observed colors into equations calibrated using standard SSP models to retrieve an approximate mass-to-light in a given photometric band.

\begin{figure}
    \centering
    \includegraphics[width=8cm]{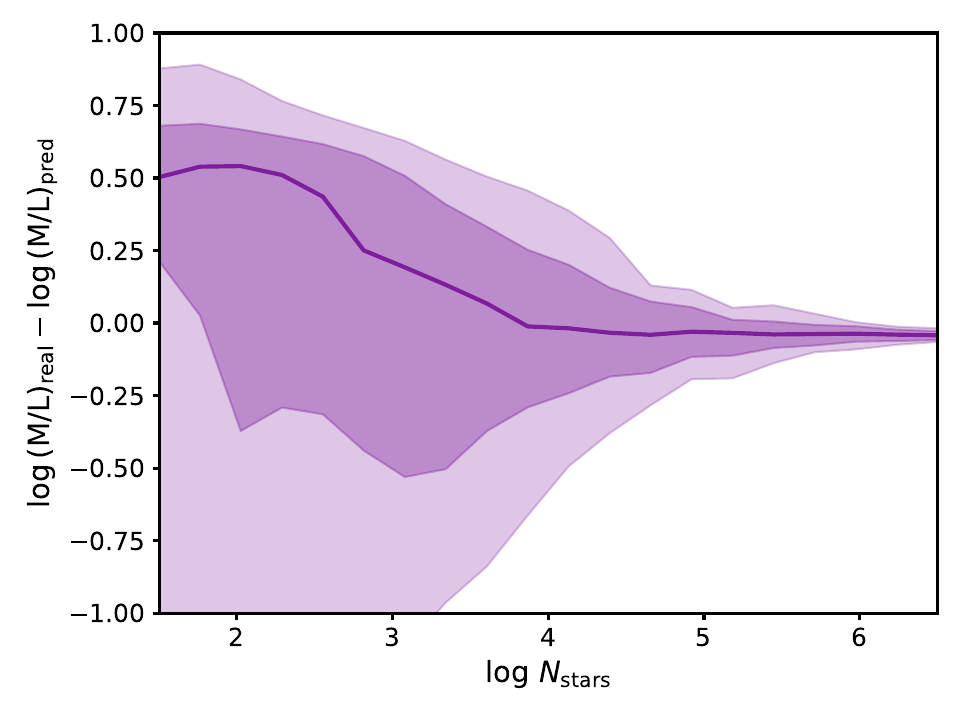}
    \caption{Color-based mass-to-light ratio estimates. The (logarithmic) ratio between the intrinsic mass-to-light ratio and that estimated using the color relations of \citet{Roediger15}, computed as a function of $N_\mathrm{stars}$ for a 10 Gyr and solar metallicity FASTAR semi-resolved population. As in the figures above, shaded areas correspond to the 10/90 and 1/99 percentiles of the measured distribution while the solid purple line corresponds to the median.}
    \label{fig:ml_pred}
 \end{figure}

Following this approach, Fig.~\ref{fig:ml_pred} compares the $r$-band mass-to-light ratios estimated from the $(g - r)$ calibration of \citet{Roediger15} with the intrinsic values measured from the FASTAR semi-resolved predictions. As before, we show the ratio between the predicted and intrinsic mass-to-light values as a function of the number of stars. The logarithmic scale on the vertical axis therefore represents the systematic error that would be incurred when using color vs mass-to-light calibrations derived from fully integrated models to estimate the stellar mass of semi-resolved populations. It is clear that color-based estimates of the mass-to-light ratio can lead to significant discrepancies between the predicted and true stellar masses of semi-resolved populations, with differences of up to $\sim$1 dex in the recovered stellar mass when the number of stars falls below $\sim10^4$, which has evident implications in, for example, measured scaling relations. Yet, the fact that empirical calibrations retain some information in the semi-resolved regime is likely due to the fact that they are ultimately sensitive to the color-luminosity-mass relation of individual stars, which are the base of both semi-resolved and fully sampled SSP models. Both Figs.~\ref{fig:ml} and~\ref{fig:ml_pred} highlight a fundamental limitation inherent to the semi-resolved regime: the number of stars is a critical parameter in determining the properties of the emitted flux, yet, its determination poses a nontrivial observational challenge. 

\section{Semi-resolved observations} \label{sec:implications}

FASTAR offers the possibility of modeling the complexity of semi-resolved stellar populations within the same framework as standard evolutionary synthesis models assuming a fully sampled IMF. As noted above, however, it is not trivial to observationally determine whether all the evolutionary phases across the IMF are sufficiently well samples in a given observation. Currently, the study of the low-surface brightness is Universe, arguably, one of the most evident situations where the stochastic sampling of the IMF might have relevant impact.

\begin{figure}
    \centering
    \includegraphics[width=8cm]{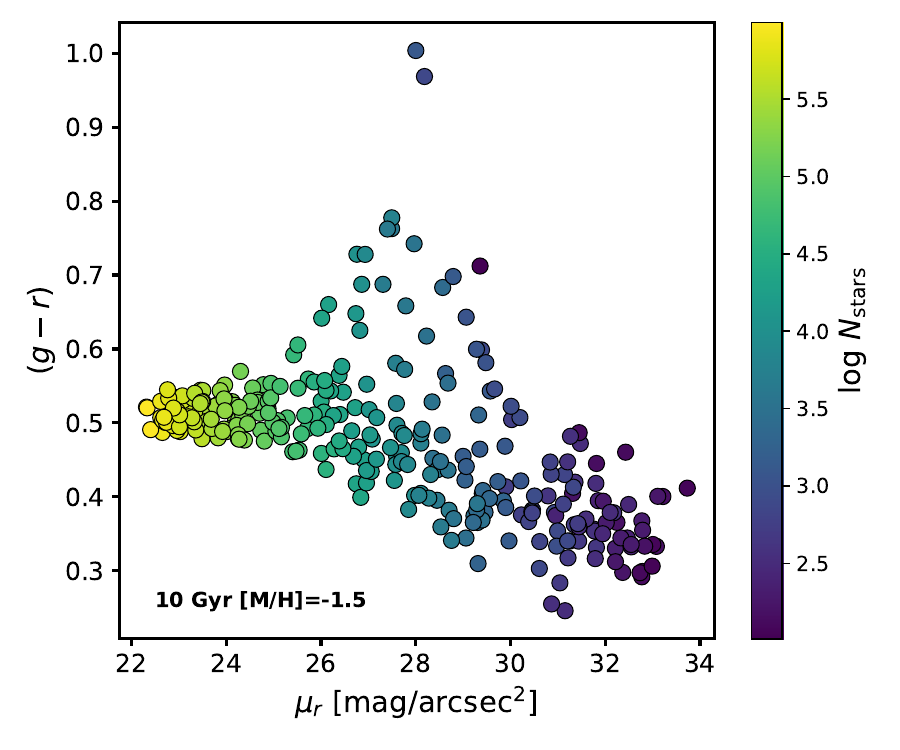}
    \caption{Colors at the Virgo distance. Colored symbols represent 300 $(g-r)$ FASTAR semi-resolved predictions for an old (10 Gyr) and metal-poor population, observed at a distance of 15 Mpc through a 1 arcsec$^2$ resolution element, representing the typical properties expected at low surface brightnesses in relatively nearby galaxies. These random FASTAR $(g-r)$ realizations are shown with varying number of stars (from left to right, as indicated by the color bar), as a function of the predicted $r$-band surface brightness.}
    \label{fig:near}
 \end{figure}

In Fig.~\ref{fig:near} we make use of the unique features of FASTAR to predict the effect of stochastic observations at different $r$-band surface brightness levels. Specifically, we show a series of arbitrary realizations of the $(g-r)$ color of a semi-resolved old and metal-poor population (10 Gyr, [M/H]$=-1.5$, as expected for the outskirts of a massive galaxy or in an extended low-mass object) vary with surface brightness. This surface brightness was calculated using the measured $r$-band absolute magnitude of each model realization, assuming that the population is at a typical distance of 15 Mpc (i.e. that of the Virgo cluster) and that the resolution element is 1~arcsec$^2$ (i.e. average seeing).

At this distance and according to Fig.~\ref{fig:near}, the $\log N_\mathrm{stars}\sim10^{5}$ threshold where the stochasticity of semi-resolved populations starts to dominate occurs at $\mu_r \sim 25$ mag arcsec$^{-2}$, well within our current technical capabilities \citep[e.g.,][]{Golini24,Khim25,Li25,Euclid}. For $\mu_r \sim 28$ there is a sudden increase in the predicted $(g-r)$ values, interpreted as the stochastic fluctuations introduced by the appearance of (red) giant stars. This is a natural prediction of semi-resolved models and should be observable even with the color uncertainty expected at this low-surface brightness level. Evidently, the exact location of this peak depends on the distance to the object and on the physical scale of the resolution element of the observations. Towards fainter magnitudes, there is an evident bias towards blue $(g-r)$ as already shown in Fig.~\ref{fig:color}.

\begin{figure}
    \centering
    \includegraphics[width=8cm]{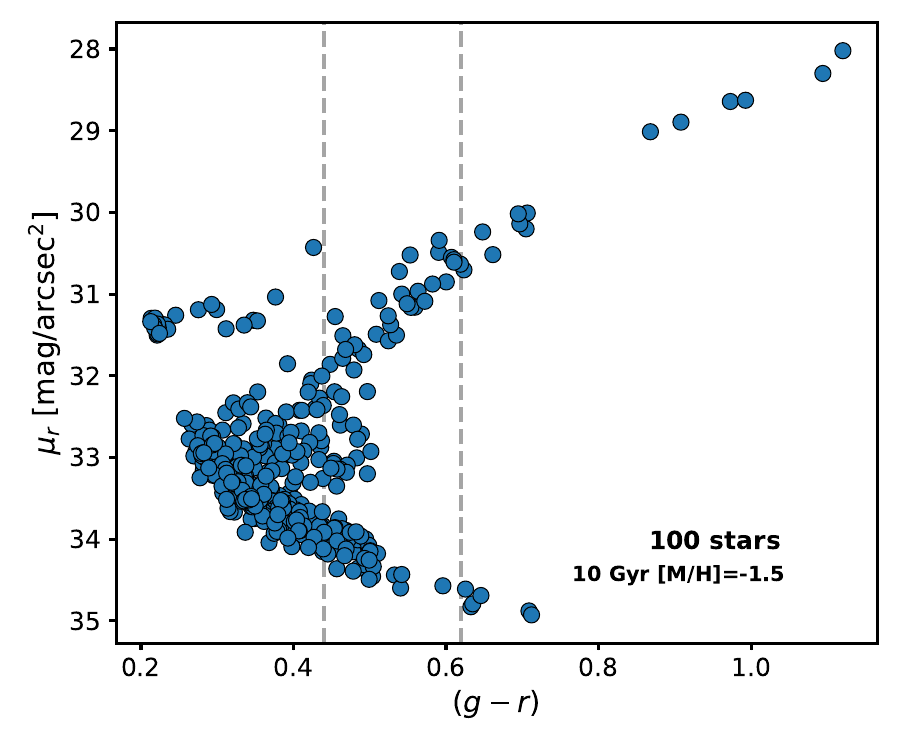}
    \includegraphics[width=8cm]{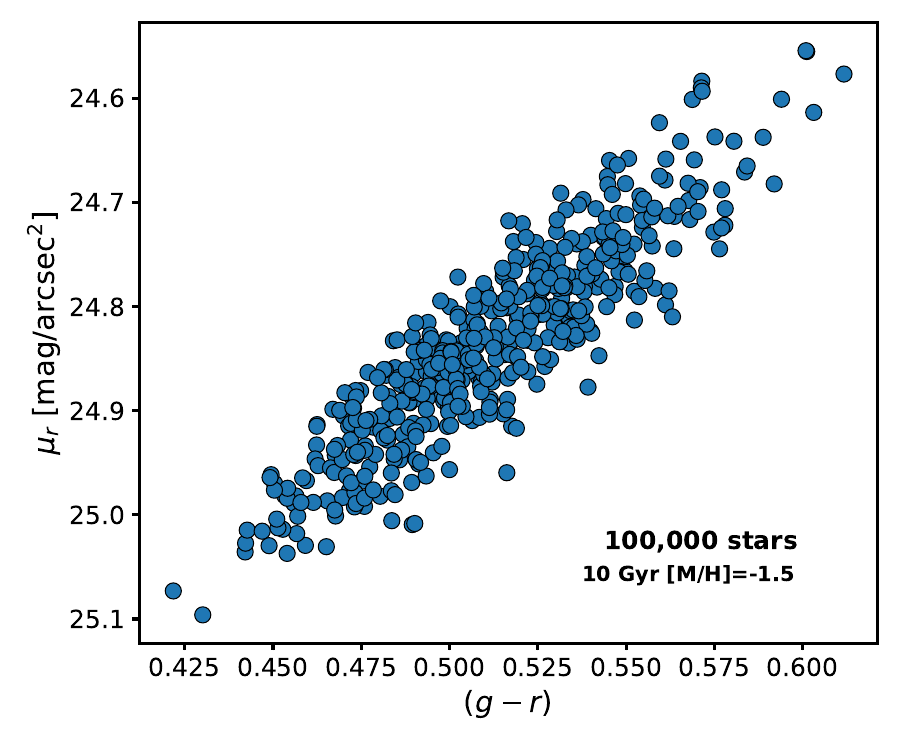}
    \caption{Isochrone mapping. Both panels represent the predicted $(g-r)$ color and $\mu_r$ surface brightness for the same old and metal-poor population of Fig.~\ref{fig:near}, observed as well with a 1 arcsec pixel scale at 15 Mpc. Each point is a different FASTAR realization with the same age, metallicity, IMF and number of stars. In the top panel, FASTAR predictions are shown for a semi-resolved population of 100 stars. Because of the sparse sampling of the IMF and thus of the different stellar evolutionary phases, the color and brightness fluctuations effectively map the underlying isochrone used to construct the model. In the bottom panel, FASTAR predictions are shown for the cases of having 100,000 stars. Measured surface brightnesses are therefore higher than in the upper panel, but exhibiting milder variations driven by red giant stars. Dashed vertical lines in the top panel illustrate the color range probed in the bottom one.}
    \label{fig:dance}
 \end{figure}

It is worth emphasizing that the correlation between scatter, observed colors, and brightness maps results from the prevalence of different stellar evolutionary phases along the isochrone as $N_\mathrm{stars}$ changes. To further illustrate this relation between isochrones (and thus stellar evolution theory) and colors, Fig.~\ref{fig:dance} shows, for two different $N_\mathrm{stars}$, how the predicted colors scale with the surface brightness level of the population. As in the figure above, color predictions in Fig.~\ref{fig:dance}  correspond to a 10 Gyr old and [M/H]$=-1.5$ stellar population.

The top panel of Fig.~\ref{fig:dance} is particularly revealing. For $N_\mathrm{stars}=100$, the color-surface brightness plane reproduces the shape of the underlying isochrone and most of its evolutionary features. As the number of stars increases, $N_\mathrm{stars}=100,000$ in the bottom panel of Fig.~\ref{fig:dance}, the variability decreases and becomes entirely dominated by fluctuations in the distribution of giant stars. Vertical dashed lines in the top panel indicate the color range probed by the bottom one, demonstrating that, in the latter, fluctuations are driven  by the stochastic sampling of giant stars. In this line, redder colors than the (g-r) shown here tend to be more sensitive to the brighter (and cooler) giant population, while fluctuations in bluer colors probe hotter evolutionary phases. Note also how, in practice, Fig.~\ref{fig:dance} corresponds approximately (given that $N_\mathrm{stars} \propto \mu_r$) to a vertical slice of the scatter in Fig.~\ref{fig:near}.

\begin{figure}
    \centering
    \includegraphics[width=8cm]{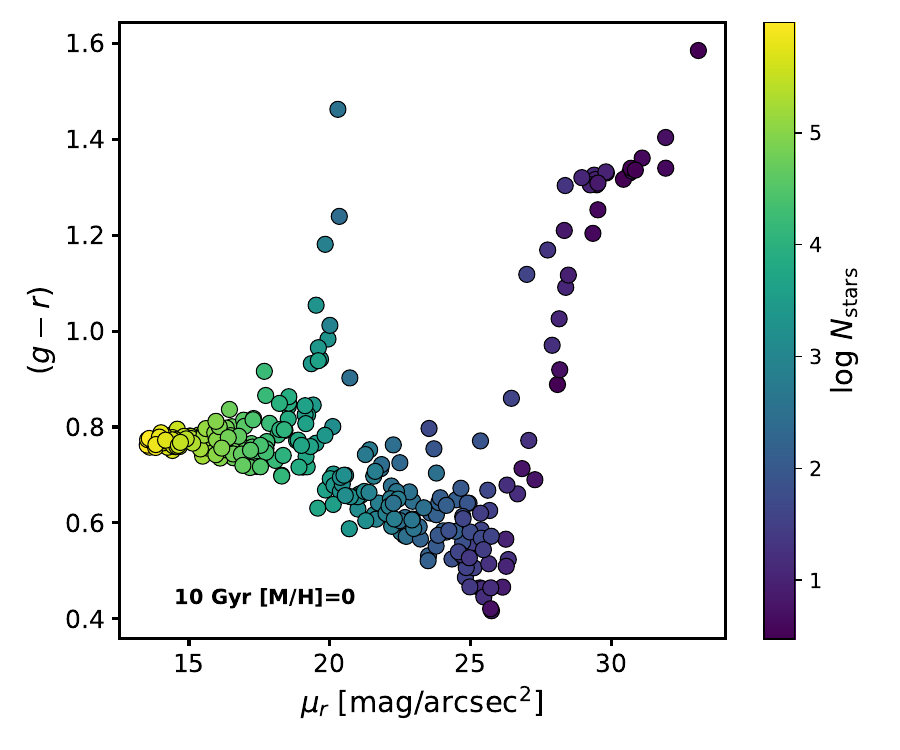}
    \caption{Predictions for 40-meter-class telescopes. Semi-resolved $(g-r)$ color predictions as a function of $\mu_r$ for an old (10 Gyr) and solar metallicity population at a distance of 100 Mpc observed projected into a 5 milli-arcsec per pixel scale, emulating a typical stellar population as observed through a 40-meter-class telescope. The effect of the stochastic sampling of the IMF becomes evident at surface brightnesses as high as $\mu_r\sim19$ mag arcsec$^{-2}$. Symbols are color-coded according to the number of stars in the semi-resolved FASTAR stellar population model.}
    \label{fig:elt}
 \end{figure}

While Fig.~\ref{fig:near} illustrates the importance that semi-resolved stellar populations already have in the analysis of low-surface brightness features, the upcoming generation of 40-meter-class telescopes possess a much greater challenge for fully sampled evolutionary stellar population models. With milli-arcsecond projected scales, the stochastic nature of semi-resolved populations will be noticeable at much higher surface brightnesses. Figure~\ref{fig:elt} represents, the $(g-r)$ color of an average old (10 Gyr) and solar metallicity population observed at 100 Mpc (i.e. the Coma cluster) through a 5 milli-arcsecond scale instrument.

Interpreting the $(g-r)$ trend and scatter in terms of the relation between IMF and stellar evolution sampling is rather evident from Fig.~\ref{fig:elt}. For very low $N_\mathrm{stars}$ the synthesized FASTAR model is dominated by the red colors of very low-mass stars \footnote{In this very low $N_\mathrm{stars}$ regime, however, the use of evolutionary stellar population models may not be the most convenient approach.}. Then, with increasing $N_\mathrm{stars}$, the color becomes bluer as the main sequence turn-off becomes more heavily sampled, reaching a minimum at around $\mu_r\sim$ 25 mag arcsec$^{-2}$. The emergence of giant stars clearly appears, in this case, at $\mu_r\sim$ 19 mag arcsec$^{-2}$. Once all stellar masses and thus evolutionary phases average out, the FASTAR $(g-r)$ converges towards the expected value of a fully sampled SSP model of 10 Gyr and solar metallicity. For a formal characterization of these biases we refer the reader to \citet{Miguel03}.

The most striking difference between Fig.~\ref{fig:near} and Fig.~\ref{fig:elt} is the surface brightness level at which stochasticity emerges. In a nearby galaxy observed with a typical seeing (Fig.~\ref{fig:near}), semi-resolved modeling becomes necessary for $\mu_r\gtrsim$ 25 mag arcsec$^{-2}$. However, even at 100 Mpc, a galaxy observed with milli-arcsecond resolution (Fig.~\ref{fig:elt}) will show signs of stochasticity in the observed data at surface brightnesses as high as $\mu_r\sim$ 19 mag arcsec$^{-2}$. For reference, only the innermost few arcseconds of the most massive galaxies in the Coma cluster reach high enough surface brightnesses \citep[e.g.,][]{Strom78} to be analyzed with fully sampled SSP models. In practice, in the era of 40-meter telescopes, Fig.~\ref{fig:elt} demonstrates that precise stellar population analyses in the local Universe will inevitably require the implementation of semi-resolved models.

\section{Stochastic predictions: Blessing or curse} \label{sec:deal}

In the sections above we have shown how FASTAR semi-resolved predictions are stochastic by nature. This is a fundamental difference with respect to standard evolutionary stellar population model predictions where the integral synthesis described by Eq.~\ref{eq:1} uniquely define the model output. This synthesis uniqueness is what justifies the use of SSP models and derived quantities such as line-strengths indices and colors to translate the spectro-photometric properties of galaxies into physically meaningful quantities.

The scatter in the semi-resolved predictions is not meaningless as it probes the biased sampling of specific stellar evolutionary phases and therefore encodes valuable information about the underlying stellar population content. In fact, fully sampled SSP models can be also understood as the mean value expected from a finite distribution of stars \citep[e.g.,][]{Miguel06} whose scatter tends to zero as the number of stars increases. Therefore, beyond the mean, higher orders of these distributions can be computed as additional constraints on the stellar population content of galaxies \citep[e.g.,][]{Vazdekis20}, this being the central idea behind the scientific exploitation of surface brightness fluctuations \citep[e.g.,][]{Tonry88,Blakeslee09,Pablo21}. Furthermore, Figs.~\ref{fig:near}, \ref{fig:dance}, and \ref{fig:elt} demonstrate how a set of independent measurements can be used as direct tests on the isochrones physics. This is a unique advantage of semi-resolved stellar populations, as they enable calibrating our knowledge on stellar evolution beyond the limited environment of the Milky Way.

Dealing with semi-resolved observations requires at the same time dedicated analysis strategies. Figures~\ref{fig:index_grid} and \ref{fig:color_index} showcase how standard observables such as colors and line-strength indices can present strong deviations compared to fully sampled SSP predictions \citep[e.g.,][]{Miguel08}. This at first sight may appear counterintuitive since averaging a large number of measurements does not lead to the average SSP value, but is simply due to the fact that equivalent widths, magnitudes, colors, mass-to-light ratios etc. are all  nonlinear operations over the emitted spectra. 

It is important to emphasize that FASTAR predictions are designed to model actual observations. The proposed random sampling of the IMF emulates, for example, the situation in which independent regions of the same globular cluster are observed in a semi-resolved manner. Even if all stars in the globular cluster share the same age and chemical composition and follow the same underlying IMF, observations of different regions will correspond to independent realizations of the IMF with a finite number of stars. However, because these realizations are drawn from a common IMF, combining multiple independent regions into a single measurement naturally converges toward the fully sampled IMF limit, i.e., the integral predictions of standard SSP models. Semi-resolved FASTAR predictions therefore constitute a natural extension of traditional evolutionary stellar population models and can be used in a similar way to fit the spectro-photometric properties of external galaxies.

Does this have real-world implications? Currently, the most evident scientific case where semi-resolved populations are at play is in the analysis of low-surface brightness features. In this regime, any stellar population signal is typically below the observational noise in the data and therefore averages (e.g. over elliptical apertures) become mandatory. However, the way in which averages are computed can have important consequences. For example, calculating an average $(g-r)$ color as the mean of the all the individual $(g-r)$ colors of every pixel will lead to the biases show in Figs.~\ref{fig:index_grid} and \ref{fig:color_index}, with the subsequent advantages and disadvantages. On the contrary, an average color measured by first averaging counts in the $g$ and $r$ bands and then calculating their magnitude difference will result, if the number of pixels is sufficiently large, into an fully sampled, SSP-equivalent measurement (assuming a linear behavior of the detector and data reduction). 

The analysis of low-surface-brightness populations presents additional challenges. Specifically, in relatively shallow observations, different evolutionary phases are expected to exhibit different signal-to-noise ratios. In particular, low-mass stars will typically contribute to the integrated properties with noisier spectra than more luminous giants. This systematic dependence of the signal-to-noise ratio on stellar luminosity may introduce additional biases beyond those discussed in the sections above, where noise was not considered. This a potential issue affecting of both semi-resolved and standard SSP models. The formalism behind the synthesis of FASTAR predictions allows for a consistent treatment of this effect. In particular, individual stellar spectra can be combined while explicitly accounting for different noise levels. Although this feature is not implemented by default, the open-source nature of FASTAR makes it straightforward to incorporate.

While the joint analysis of a large number of independent measurements offers a viable way forward to extract information from semi-resolved observations, being able to analyze individual spectro-photometric measurements remains, for obvious reasons, a desirable goal. The stochastic nature of semi-resolved observations posses, however, an evident challenge. Standard inversion algorithms cannot deal with the inherent variability of semi-resolved models, as the same input parameters (age, metallicity, IMF and number of stars in the case of FASTAR) can result in drastically different model outputs. In this context, simulation based inference approaches offer a clear and formal way forward to deal with such intractable likelihoods \citep[e.g.,][]{Hahn22,Eirini23,Eirini24,Eirini25,Patricia24,Patricia25} and the efficient synthesis of semi-resolved FASTAR models is particularly well-suited for these forward modeling tasks, as we will exemplify in upcoming papers. Complementary, FASTAR also offers the framework to combine independent measurements of the same stellar population in order to derive tighter observational constraints.

This interpretation of semi-resolved observations must also consider the possibility of composite stellar populations. It can be argued that as the physical scale probed within a galaxy decreases, and consequently $N_\mathrm{stars}$ decreases, the stellar population properties become progressively simpler. For example, the integrated spectrum of a spiral galaxy containing both a bulge and a disc is generally more complex than the spectra of each component separately. Observations at smaller spatial scales therefore tend to minimize the effect of radial stellar population gradients, justifying the SSP assumption. Nevertheless, as in standard SSP models, the spectrum of a composite population remains a linear combination of the spectra of the individual subpopulations. Semi-resolved models therefore preserve the linearity that underlies most population inference techniques, allowing composite populations to be incorporated naturally within inversion frameworks in much the same way as in fitting methods based on fully sampled SSP models \citep[e.g.,][]{ppxf,Ocvirk06,bagpipes,prospector}. In practice, this means that semi-resolved modeling does not introduce a fundamentally different treatment of composite populations, but rather extends the standard SSP framework to regimes where IMF sampling becomes relevant.

Finally, a note on the term semi-resolved. Throughout this paper, we use this term to describe situations in which the number of stars per resolution element is not large enough to properly sample all stellar evolutionary phases along an isochrone, and therefore constitutes a regime in which the standard integral synthesis of evolutionary stellar population models (Eq.~\ref{eq:1}) does not strictly apply. Because different stellar evolutionary phases present prominent color differences, the boundary between both regimes depends on the exact wavelength range \citep[e.g.,][]{Miguel04}. As clearly explained in Section 1 of \citet{Conroy16}, the concept of a semi-resolved population is intended to capture the intermediate regime between observations in which all individual stars can be resolved and the opposite extreme, in which the observed spectra are effectively composed of an infinite number of stars. In many semi-resolved situations, however, one should not expect individual stars to be partially resolved, as observations will only capture their integrated flux. In short, semi-resolved evolutionary stellar population models aim to reproduce the combined flux of a relatively small number of stars with the same age and chemical composition, all observed within a single resolution element.


\section{Summary}\label{sec:summary}

FASTAR enables the modeling of semi-resolved stellar populations based on the same evolutionary principles as standard SSP models. Thanks to its computational efficiency, semi-resolved models can be computed on the fly over a wide range of ages and metallicities. The combination of empirical and theoretical stellar libraries allows for detailed spectroscopic predictions from $3,540$ to $7,400$ $\AA$ and from $2,000$ to $12,000$ $\AA$ for photometric applications. The calibration of the FASTAR models ensures that arbitrary line-strength indices, colors, magnitudes, and mass-to-light ratios can be computed. 

Beyond the description of the model synthesis, the main features of FASTAR semi-resolved predictions are as follows:

\begin{itemize}
    \item Semi-resolved predictions share the same assumptions of standard evolutionary stellar population models but naturally extend their applicability range from fully-integrated to fully-resolved observations.
    \item Stochastic variations in the model predictions are noticeable even if the number of stars is as high as $N_\mathrm{stars} \sim 10^6$. When the number of stars drops below $N_\mathrm{stars} \sim 10^{4.5}$, stochasticity becomes a dominant factor determining the properties of FASTAR semi-resolved spectra.
    \item Line-strength indices traditionally used to infer stellar population properties from integrated spectra lose much of their constraining power in the semi-resolved regime. For broader photometric measurements such as colors, the effect of stochasticity is even more pronounced. The scatter and potential biases in the line-strengths and colors of semi-resolved populations heavily depend on the adopted IMF.
    \item While the number of stars is a critical parameter in modeling semi-resolved populations, it is challenging to constrain from observations as the mass-to-light ratio in the semi-resolved regime can present large variations and systematic biases even if the underlying stellar population parameters (namely age, metallicity and IMF) remain constant. Mass-to-light ratio approximations based on color scaling relations present similar difficulties.
    \item The behavior of semi-resolved FASTAR spectra as a function of $N_\mathrm{stars}$ is driven by the incomplete sampling of different stellar evolutionary phases. Only when all these phases are properly sampled, FASTAR predictions converge towards the integral formulation. Conversely, semi-resolved observations can be used as direct tests of stellar evolution theory.
    \item FASTAR models predict that stochastic effects should be already noticeable in low-surface brightness studies. Looking ahead, semi-resolved observations are expected to be the norm for the next generation of 40-meter telescopes.
\end{itemize}

Semi-resolved stellar populations pose important technical challenges. The inherent stochasticity due to the incomplete IMF sampling implies that there are no unique predictions, rendering standard stellar population inversion algorithms unusable. Furthermore, at a deeper level, stellar atmospheres do not contain  information about the age or (evidently) IMF of a population. Therefore, from the point of view of the integrated light, these quantities are only measurable when analyzing a sufficiently large collection of coeval stars. Thus, contrary to standard fully sampled evolutionary models, in the semi-resolved regime our ability to determine ages and IMFs is inevitably conditioned by the total number of stars in our resolution element.

While the distribution of observed values (brightnesses, colors, line indices etc.) holds value information about the properties of semi-resolved populations, analysis approaches able to fit individual observations are highly desirable in order to maximize the information extracted from current and upcoming facilities. In this context, FASTAR used in combination with new techniques such as simulation-based inference frameworks \citep[e.g.,][]{Hahn22,Patricia24,Patricia25} offers a promising way forward.

\begin{acknowledgements}
We would like to thank the referee for the insightful and constructive interactions. We acknowledge support from grant PID2022-140869NB-I00 funded by the Spanish Ministry of Science and Innovation, from PID2022-136598NB-C33 funded by MCIN/AEI/10.13039/501100011033 and by “ERDF A way of making Europe”.  F.L.B. acknowledges support from INAF minigrant 1.05.23.04.01. We would like to thank Ignacio Ferreras and Ignacio Trujillo for their insightful comments during the development of FASTAR.
\end{acknowledgements}

\section*{Models availability}
Documentation and examples on how to use FASTAR can be found at the project's website
\url{https://fastar.readthedocs.io}

The code is freely available here
\url{https://github.com/inavarro/fastar}

\bibliographystyle{aa}  
\bibliography{fastar_semi} 

@Article{ppxf,
  Title                    = {{Parametric Recovery of Line-of-Sight Velocity Distributions from Absorption-Line Spectra of Galaxies via Penalized Likelihood}},
  Author                   = {{Cappellari}, M. and {Emsellem}, E.},
  Journal                  = {\pasp},
  Year                     = {2004},

  Month                    = feb,
  Pages                    = {138-147},
  Volume                   = {116},
  Adsurl                   = {http://adsabs.harvard.edu/abs/2004PASP..116..138C},
  Doi                      = {10.1086/381875},
  Eprint                   = {astro-ph/0312201}
}

@Article{mw,
  Title                    = {{On the variation of the initial mass function}},
  Author                   = {{Kroupa}, P.},
  Journal                  = {\mnras},
  Year                     = {2001},

  Month                    = apr,
  Pages                    = {231-246},
  Volume                   = {322},
  Adsurl                   = {http://adsabs.harvard.edu/abs/2001MNRAS.322..231K},
  Doi                      = {10.1046/j.1365-8711.2001.04022.x},
  Eprint                   = {astro-ph/0009005}
}

@Article{Spiniello2013,
  Title                    = {{The stellar IMF in early-type galaxies from a non-degenerate set of optical line indices}},
  Author                   = {{Spiniello}, C. and {Trager}, S. and {Koopmans}, L.~V.~E. and {Conroy}, C.},
  Journal                  = {\mnras},
  Year                     = {2014},

  Month                    = feb,
  Pages                    = {1483-1499},
  Volume                   = {438},
  Adsurl                   = {http://adsabs.harvard.edu/abs/2014MNRAS.438.1483S},
  Archiveprefix            = {arXiv},
  Doi                      = {10.1093/mnras/stt2282},
  Eprint                   = {1305.2873},
  Primaryclass             = {astro-ph.CO}
}

@Article{TMB:03,
  Title                    = {{Stellar population models of Lick indices with variable element abundance ratios}},
  Author                   = {{Thomas}, D. and {Maraston}, C. and {Bender}, R.},
  Journal                  = {\mnras},
  Year                     = {2003},

  Month                    = mar,
  Pages                    = {897-911},
  Volume                   = {339},
  Adsurl                   = {http://adsabs.harvard.edu/abs/2003MNRAS.339..897T},
  Doi                      = {10.1046/j.1365-8711.2003.06248.x},
  Eprint                   = {astro-ph/0209250}
}

@Article{trager,
  Title                    = {{Old Stellar Populations. VI. Absorption-Line Spectra of Galaxy Nuclei and Globular Clusters}},
  Author                   = {{Trager}, S.~C. and {Worthey}, G. and {Faber}, S.~M. and {Burstein}, D. and {Gonzalez}, J.~J.},
  Journal                  = {\apjs},
  Year                     = {1998},

  Month                    = may,
  Pages                    = {1},
  Volume                   = {116},
  Adsurl                   = {http://adsabs.harvard.edu/abs/1998ApJS..116....1T},
  Doi                      = {10.1086/313099},
  Eprint                   = {astro-ph/9712258}
}

@Article{vazdekis96,
  Title                    = {{A New Chemo-evolutionary Population Synthesis Model for Early-Type Galaxies. I. Theoretical Basis}},
  Author                   = {{Vazdekis}, A. and {Casuso}, E. and {Peletier}, R.~F. and {Beckman}, J.~E. },
  Journal                  = {\apjs},
  Year                     = {1996},

  Month                    = oct,
  Pages                    = {307},
  Volume                   = {106},
  Adsurl                   = {http://adsabs.harvard.edu/abs/1996ApJS..106..307V},
  Doi                      = {10.1086/192340},
  Eprint                   = {astro-ph/9605112}
}

@Article{miles,
  Title                    = {{Evolutionary stellar population synthesis with MILES - I. The base models and a new line index system}},
  Author                   = {{Vazdekis}, A. and {S{\'a}nchez-Bl{\'a}zquez}, P. and {Falc{\'o}n-Barroso}, J. and {Cenarro}, A.~J. and {Beasley}, M.~A. and {Cardiel}, N. and {Gorgas}, J. and {Peletier}, R.~F.},
  Journal                  = {\mnras},
  Year                     = {2010},

  Month                    = jun,
  Pages                    = {1639-1671},
  Volume                   = {404},
  Adsurl                   = {http://adsabs.harvard.edu/abs/2010MNRAS.404.1639V},
  Archiveprefix            = {arXiv},
  Doi                      = {10.1111/j.1365-2966.2010.16407.x},
  Eprint                   = {1004.4439},
  Primaryclass             = {astro-ph.CO}
}

@Article{Chabrier,
  author   = {{Chabrier}, G.},
  title    = {{Galactic Stellar and Substellar Initial Mass Function}},
  journal  = {\pasp},
  year     = {2003},
  volume   = {115},
  pages    = {763-795},
  month    = jul,
  adsurl   = {http://adsabs.harvard.edu/abs/2003PASP..115..763C},
  doi      = {10.1086/376392},
  eprint   = {astro-ph/0304382},
}

@Article{Conroy12,
  author        = {{Conroy}, C. and {van Dokkum}, P.~G.},
  title         = {{The Stellar Initial Mass Function in Early-type Galaxies From Absorption Line Spectroscopy. II. Results}},
  journal       = {\apj},
  year          = {2012},
  volume        = {760},
  pages         = {71},
  month         = nov,
  adsurl        = {http://adsabs.harvard.edu/abs/2012ApJ...760...71C},
  archiveprefix = {arXiv},
  doi           = {10.1088/0004-637X/760/1/71},
  eid           = {71},
  eprint        = {1205.6473},
}

@Article{Blakeslee09,
  author        = {{Blakeslee}, J.~P. and {Jord{\'a}n}, A. and {Mei}, S. and {C{\^o}t{\'e}}, P. and {Ferrarese}, L. and {Infante}, L. and {Peng}, E.~W. and {Tonry}, J.~L. and {West}, M.~J.},
  title         = {{The ACS Fornax Cluster Survey. V. Measurement and Recalibration of Surface Brightness Fluctuations and a Precise Value of the Fornax-Virgo Relative Distance}},
  journal       = {\apj},
  year          = {2009},
  volume        = {694},
  pages         = {556-572},
  month         = mar,
  adsurl        = {http://adsabs.harvard.edu/abs/2009ApJ...694..556B},
  archiveprefix = {arXiv},
  doi           = {10.1088/0004-637X/694/1/556},
  eprint        = {0901.1138},
  primaryclass  = {astro-ph.CO},
}

@Article{Vazdekis15,
  author        = {{Vazdekis}, A. and {Coelho}, P. and {Cassisi}, S. and {Ricciardelli}, E. and {Falc{\'o}n-Barroso}, J. and {S{\'a}nchez-Bl{\'a}zquez}, P. and {La Barbera}, F. and {Beasley}, M.~A. and {Pietrinferni}, A.},
  title         = {{Evolutionary stellar population synthesis with MILES - II. Scaled-solar and {$\alpha$}-enhanced models}},
  journal       = {\mnras},
  year          = {2015},
  volume        = {449},
  pages         = {1177-1214},
  month         = may,
  adsurl        = {http://adsabs.harvard.edu/abs/2015MNRAS.449.1177V},
  archiveprefix = {arXiv},
  doi           = {10.1093/mnras/stv151},
  eprint        = {1504.08032},
}

@Article{Pat06,
  author   = {{S{\'a}nchez-Bl{\'a}zquez}, P. and {Peletier}, R.~F. and {Jim{\'e}nez-Vicente}, J. and {Cardiel}, N. and {Cenarro}, A.~J. and {Falc{\'o}n-Barroso}, J. and {Gorgas}, J. and {Selam}, S. and {Vazdekis}, A.},
  title    = {{Medium-resolution Isaac Newton Telescope library of empirical spectra}},
  journal  = {\mnras},
  year     = {2006},
  volume   = {371},
  pages    = {703-718},
  month    = sep,
  adsurl   = {http://adsabs.harvard.edu/abs/2006MNRAS.371..703S},
  doi      = {10.1111/j.1365-2966.2006.10699.x},
  eprint   = {astro-ph/0607009},
}

@Article{Jesus11,
  author        = {{Falc{\'o}n-Barroso}, J. and {S{\'a}nchez-Bl{\'a}zquez}, P. and {Vazdekis}, A. and {Ricciardelli}, E. and {Cardiel}, N. and {Cenarro}, A.~J. and {Gorgas}, J. and {Peletier}, R.~F.},
  title         = {{An updated MILES stellar library and stellar population models}},
  journal       = {\aap},
  year          = {2011},
  volume        = {532},
  pages         = {A95},
  month         = aug,
  adsurl        = {http://adsabs.harvard.edu/abs/2011A%26A...532A..95F},
  archiveprefix = {arXiv},
  doi           = {10.1051/0004-6361/201116842},
  eid           = {A95},
  eprint        = {1107.2303},
}

@Article{Schiavon07,
  author   = {{Schiavon}, R.~P.},
  title    = {{Population Synthesis in the Blue. IV. Accurate Model Predictions for Lick Indices and UBV Colors in Single Stellar Populations}},
  journal  = {\apjs},
  year     = {2007},
  volume   = {171},
  pages    = {146-205},
  month    = jul,
  adsurl   = {http://adsabs.harvard.edu/abs/2007ApJS..171..146S},
  doi      = {10.1086/511753},
  eprint   = {astro-ph/0611464},
}

@Article{Worthey94,
  author   = {{Worthey}, G.},
  title    = {{Comprehensive stellar population models and the disentanglement of age and metallicity effects}},
  journal  = {\apjs},
  year     = {1994},
  volume   = {95},
  pages    = {107-149},
  month    = nov,
  adsurl   = {http://adsabs.harvard.edu/abs/1994ApJS...95..107W},
  doi      = {10.1086/192096},
}

@Article{Ocvirk06,
  author   = {{Ocvirk}, P. and {Pichon}, C. and {Lan{\c c}on}, A. and {Thi{\'e}baut}, E.},
  title    = {{STECKMAP: STEllar Content and Kinematics from high resolution galactic spectra via Maximum A Posteriori}},
  journal  = {\mnras},
  year     = {2006},
  volume   = {365},
  pages    = {74-84},
  month    = jan,
  adsurl   = {http://adsabs.harvard.edu/abs/2006MNRAS.365...74O},
  doi      = {10.1111/j.1365-2966.2005.09323.x},
  eprint   = {astro-ph/0507002},
}

@Article{Worthey92,
  author   = {{Worthey}, G. and {Faber}, S.~M. and {Gonzalez}, J.~J.},
  title    = {{MG and Fe absorption features in elliptical galaxies}},
  journal  = {\apj},
  year     = {1992},
  volume   = {398},
  pages    = {69-73},
  month    = oct,
  adsurl   = {http://adsabs.harvard.edu/abs/1992ApJ...398...69W},
  doi      = {10.1086/171836},
}

@Article{Thomas05,
  author   = {{Thomas}, D. and {Maraston}, C. and {Bender}, R. and {Mendes de Oliveira}, C.},
  title    = {{The Epochs of Early-Type Galaxy Formation as a Function of Environment}},
  journal  = {\apj},
  year     = {2005},
  volume   = {621},
  pages    = {673-694},
  month    = mar,
  adsurl   = {http://adsabs.harvard.edu/abs/2005ApJ...621..673T},
  doi      = {10.1086/426932},
  eprint   = {astro-ph/0410209},
}

@Article{Kuntschner10,
  author        = {{Kuntschner}, H. and {Emsellem}, E. and {Bacon}, R. and {Cappellari}, M. and {Davies}, R.~L. and {de Zeeuw}, P.~T. and {Falc{\'o}n-Barroso}, J. and {Krajnovi{\'c}}, D. and {McDermid}, R.~M. and {Peletier}, R.~F. and {Sarzi}, M. and {Shapiro}, K.~L. and {van den Bosch}, R.~C.~E. and {van de Ven}, G.},
  title         = {{The SAURON project - XVII. Stellar population analysis of the absorption line strength maps of 48 early-type galaxies}},
  journal       = {\mnras},
  year          = {2010},
  volume        = {408},
  pages         = {97-132},
  month         = oct,
  adsurl        = {http://adsabs.harvard.edu/abs/2010MNRAS.408...97K},
  archiveprefix = {arXiv},
  doi           = {10.1111/j.1365-2966.2010.17161.x},
  eprint        = {1006.1574},
}

@Article{Conroy13,
  author        = {{Conroy}, Charlie},
  title         = {{Modeling the Panchromatic Spectral Energy Distributions of Galaxies}},
  journal       = {\araa},
  year          = {2013},
  volume        = {51},
  number        = {1},
  pages         = {393-455},
  month         = aug,
  adsurl        = {https://ui.adsabs.harvard.edu/abs/2013ARA&A..51..393C},
  archiveprefix = {arXiv},
  doi           = {10.1146/annurev-astro-082812-141017},
  eprint        = {1301.7095},
  primaryclass  = {astro-ph.CO},
}

@Article{LB19,
  author        = {{La Barbera}, F. and {Vazdekis}, A. and {Ferreras}, I. and {Pasquali}, A. and {Allende Prieto}, C. and {Mart{\'\i}n-Navarro}, I. and {Aguado}, D.~S. and {de Carvalho}, R.~R. and {Rembold}, S. and {Falc{\'o}n-Barroso}, J. and {van de Ven}, G.},
  title         = {{IMF radial gradients in most massive early-type galaxies}},
  journal       = {\mnras},
  year          = {2019},
  volume        = {489},
  number        = {3},
  pages         = {4090-4110},
  month         = nov,
  adsurl        = {https://ui.adsabs.harvard.edu/abs/2019MNRAS.489.4090L},
  archiveprefix = {arXiv},
  doi           = {10.1093/mnras/stz2192},
  eprint        = {1909.01382},
  primaryclass  = {astro-ph.GA},
}

@Article{Pinna19b,
  author        = {{Pinna}, F. and {Falc{\'o}n-Barroso}, J. and {Martig}, M. and {Sarzi}, M. and {Coccato}, L. and {Iodice}, E. and {Corsini}, E.~M. and {de Zeeuw}, P.~T. and {Gadotti}, D.~A. and {Leaman}, R. and {Lyubenova}, M. and {McDermid}, R.~M. and {Minchev}, I. and {Morelli}, L. and {van de Ven}, G. and {Viaene}, S.},
  title         = {{The Fornax 3D project: Unveiling the thick disk origin in FCC 170; possible signs of accretion}},
  journal       = {\aap},
  year          = {2019},
  volume        = {623},
  pages         = {A19},
  month         = mar,
  adsurl        = {https://ui.adsabs.harvard.edu/abs/2019A&A...623A..19P},
  archiveprefix = {arXiv},
  doi           = {10.1051/0004-6361/201833193},
  eid           = {A19},
  eprint        = {1901.04310},
  primaryclass  = {astro-ph.GA},
}

@Article{Justus2020,
  author        = {{Neumann}, Justus and {Fragkoudi}, Francesca and {P{\'e}rez}, Isabel and {Gadotti}, Dimitri A. and {Falc{\'o}n-Barroso}, Jes{\'u}s and {S{\'a}nchez-Bl{\'a}zquez}, Patricia and {Bittner}, Adrian and {Husemann}, Bernd and {G{\'o}mez}, Facundo A. and {Grand}, Robert J.~J. and {Donohoe-Keyes}, Charlotte E. and {Kim}, Taehyun and {de Lorenzo-C{\'a}ceres}, Adriana and {Martig}, Marie and {M{\'e}ndez-Abreu}, Jairo and {Pakmor}, R{\"u}diger and {Seidel}, Marja K. and {van de Ven}, Glenn},
  title         = {{Stellar populations across galaxy bars in the MUSE TIMER project}},
  journal       = {\aap},
  year          = {2020},
  volume        = {637},
  pages         = {A56},
  month         = may,
  adsurl        = {https://ui.adsabs.harvard.edu/abs/2020A&A...637A..56N},
  archiveprefix = {arXiv},
  doi           = {10.1051/0004-6361/202037604},
  eid           = {A56},
  eprint        = {2003.08946},
  primaryclass  = {astro-ph.GA},
}

@Article{Tinsley76,
  author   = {{Tinsley}, B.~M. and {Gunn}, J.~E.},
  title    = {{Evolutionary synthesis of the stellar population in elliptical galaxies. I. Ingredients, broad-band colors, and infrared features.}},
  journal  = {\apj},
  year     = {1976},
  volume   = {203},
  pages    = {52-62},
  month    = jan,
  adsurl   = {https://ui.adsabs.harvard.edu/abs/1976ApJ...203...52T},
  doi      = {10.1086/154046},
}

@Article{MN21,
  author        = {{Mart{\'\i}n-Navarro}, I. and {Pinna}, F. and {Coccato}, L. and {Falc{\'o}n-Barroso}, J. and {van de Ven}, G. and {Lyubenova}, M. and {Corsini}, E.~M. and {Fahrion}, K. and {Gadotti}, D.~A. and {Iodice}, E. and {McDermid}, R.~M. and {Poci}, A. and {Sarzi}, M. and {Spriggs}, T.~W. and {Viaene}, S. and {de Zeeuw}, P.~T. and {Zhu}, L.},
  title         = {{Fornax 3D project: Assessing the diversity of IMF and stellar population maps within the Fornax Cluster}},
  journal       = {\aap},
  year          = {2021},
  volume        = {654},
  pages         = {A59},
  month         = oct,
  adsurl        = {https://ui.adsabs.harvard.edu/abs/2021A&A...654A..59M},
  archiveprefix = {arXiv},
  doi           = {10.1051/0004-6361/202141348},
  eid           = {A59},
  eprint        = {2107.14243},
  primaryclass  = {astro-ph.GA},
}

@Article{Cervantes,
  author        = {{Cervantes}, J.~L. and {Vazdekis}, A.},
  title         = {{An optimized H{\ensuremath{\beta}} index for disentangling stellar population ages}},
  journal       = {\mnras},
  year          = {2009},
  volume        = {392},
  number        = {2},
  pages         = {691-704},
  month         = jan,
  adsurl        = {https://ui.adsabs.harvard.edu/abs/2009MNRAS.392..691C},
  archiveprefix = {arXiv},
  doi           = {10.1111/j.1365-2966.2008.14079.x},
  eprint        = {0810.3240},
  primaryclass  = {astro-ph},
}

@Article{Worthey94b,
  author   = {{Worthey}, Guy and {Faber}, S.~M. and {Gonzalez}, J. Jesus and {Burstein}, D.},
  title    = {{Old Stellar Populations. V. Absorption Feature Indices for the Complete Lick/IDS Sample of Stars}},
  journal  = {\apjs},
  year     = {1994},
  volume   = {94},
  pages    = {687},
  month    = oct,
  adsurl   = {https://ui.adsabs.harvard.edu/abs/1994ApJS...94..687W},
  doi      = {10.1086/192087},
}

@Article{Leitherer99,
  author        = {{Leitherer}, Claus and {Schaerer}, Daniel and {Goldader}, Jeffrey D. and {Delgado}, Rosa M. Gonz{\'a}lez and {Robert}, Carmelle and {Kune}, Denis Foo and {de Mello}, Du{\'\i}lia F. and {Devost}, Daniel and {Heckman}, Timothy M.},
  title         = {{Starburst99: Synthesis Models for Galaxies with Active Star Formation}},
  journal       = {\apjs},
  year          = {1999},
  volume        = {123},
  number        = {1},
  pages         = {3-40},
  month         = jul,
  adsurl        = {https://ui.adsabs.harvard.edu/abs/1999ApJS..123....3L},
  archiveprefix = {arXiv},
  doi           = {10.1086/313233},
  eprint        = {astro-ph/9902334},
  primaryclass  = {astro-ph},
}

@Article{Bittner20,
  author        = {{Bittner}, Adrian and {S{\'a}nchez-Bl{\'a}zquez}, Patricia and {Gadotti}, Dimitri A. and {Neumann}, Justus and {Fragkoudi}, Francesca and {Coelho}, Paula and {de Lorenzo-C{\'a}ceres}, Adriana and {Falc{\'o}n-Barroso}, Jes{\'u}s and {Kim}, Taehyun and {Leaman}, Ryan and {Mart{\'\i}n-Navarro}, Ignacio and {M{\'e}ndez-Abreu}, Jairo and {P{\'e}rez}, Isabel and {Querejeta}, Miguel and {Seidel}, Marja K. and {van de Ven}, Glenn},
  title         = {{Inside-out formation of nuclear discs and the absence of old central spheroids in barred galaxies of the TIMER survey}},
  journal       = {\aap},
  year          = {2020},
  volume        = {643},
  pages         = {A65},
  month         = nov,
  adsurl        = {https://ui.adsabs.harvard.edu/abs/2020A&A...643A..65B},
  archiveprefix = {arXiv},
  doi           = {10.1051/0004-6361/202038450},
  eid           = {A65},
  eprint        = {2009.01856},
  primaryclass  = {astro-ph.GA},
}

@Article{prospector,
  author        = {{Johnson}, Benjamin D. and {Leja}, Joel and {Conroy}, Charlie and {Speagle}, Joshua S.},
  title         = {{Stellar Population Inference with Prospector}},
  journal       = {\apjs},
  year          = {2021},
  volume        = {254},
  number        = {2},
  pages         = {22},
  month         = jun,
  adsurl        = {https://ui.adsabs.harvard.edu/abs/2021ApJS..254...22J},
  archiveprefix = {arXiv},
  doi           = {10.3847/1538-4365/abef67},
  eid           = {22},
  eprint        = {2012.01426},
  primaryclass  = {astro-ph.GA},
}

@Article{Faber72,
  author  = {{Faber}, S.~M.},
  title   = {{Quadratic programming applied to the problem of galaxy population synthesis.}},
  journal = {\aap},
  year    = {1972},
  volume  = {20},
  pages   = {361},
  month   = sep,
  adsurl  = {https://ui.adsabs.harvard.edu/abs/1972A&A....20..361F},
}

@Article{Spinrad71,
  author  = {{Spinrad}, Hyron and {Taylor}, Benjamin J.},
  title   = {{The Stellar Content of the Nuclei of Nearby Galaxies. I. M31, M32, and M81}},
  journal = {\apjs},
  year    = {1971},
  volume  = {22},
  pages   = {445},
  month   = apr,
  adsurl  = {https://ui.adsabs.harvard.edu/abs/1971ApJS...22..445S},
  doi     = {10.1086/190232},
}

@Article{Bica86,
  author   = {{Bica}, E. and {Alloin}, D.},
  title    = {{A base of star clusters for stellar population synthesis.}},
  journal  = {\aap},
  year     = {1986},
  volume   = {162},
  pages    = {21-31},
  month    = jul,
  adsurl   = {https://ui.adsabs.harvard.edu/abs/1986A&A...162...21B},
}

@Article{Pelat98,
  author   = {{Pelat}, D.},
  title    = {{Stellar population synthesis with more degrees of freedom than observables}},
  journal  = {\mnras},
  year     = {1998},
  volume   = {299},
  number   = {3},
  pages    = {877-888},
  month    = sep,
  adsurl   = {https://ui.adsabs.harvard.edu/abs/1998MNRAS.299..877P},
  doi      = {10.1046/j.1365-8711.1998.01825.x},
}

@Article{Pickles,
  author   = {{Pickles}, A.~J.},
  title    = {{Differential population synthesis of early-type galaxies. III. Synthesis results.}},
  journal  = {\apj},
  year     = {1985},
  volume   = {296},
  pages    = {340-369},
  month    = sep,
  adsurl   = {https://ui.adsabs.harvard.edu/abs/1985ApJ...296..340P},
  doi      = {10.1086/163454},
}

@Article{Schmidt91,
  author   = {{Schmidt}, Alex A. and {Copetti}, Marcus V.~F. and {Alloin}, Danielle and {Jablonka}, Pascale},
  title    = {{Population synthesis methods : discussion and tests on the solution uniqueness.}},
  journal  = {\mnras},
  year     = {1991},
  volume   = {249},
  pages    = {766},
  month    = apr,
  adsurl   = {https://ui.adsabs.harvard.edu/abs/1991MNRAS.249..766S},
  doi      = {10.1093/mnras/249.4.766},
}

@Article{bc03,
  author        = {{Bruzual}, G. and {Charlot}, S.},
  title         = {{Stellar population synthesis at the resolution of 2003}},
  journal       = {\mnras},
  year          = {2003},
  volume        = {344},
  number        = {4},
  pages         = {1000-1028},
  month         = oct,
  adsurl        = {https://ui.adsabs.harvard.edu/abs/2003MNRAS.344.1000B},
  archiveprefix = {arXiv},
  doi           = {10.1046/j.1365-8711.2003.06897.x},
  eprint        = {astro-ph/0309134},
  primaryclass  = {astro-ph},
}

@Article{Maraston05,
  author        = {{Maraston}, Claudia},
  title         = {{Evolutionary population synthesis: models, analysis of the ingredients and application to high-z galaxies}},
  journal       = {\mnras},
  year          = {2005},
  volume        = {362},
  number        = {3},
  pages         = {799-825},
  month         = sep,
  adsurl        = {https://ui.adsabs.harvard.edu/abs/2005MNRAS.362..799M},
  archiveprefix = {arXiv},
  doi           = {10.1111/j.1365-2966.2005.09270.x},
  eprint        = {astro-ph/0410207},
  primaryclass  = {astro-ph},
}

@Article{LVM,
  author        = {{Drory}, Niv and {Blanc}, Guillermo A. and {Kreckel}, Kathryn and {S{\'a}nchez}, Sebasti{\'a}n F. and {Mej{\'\i}a-Narv{\'a}ez}, Alfredo and {Johnston}, Evelyn J. and {Jones}, Amy M. and {Pellegrini}, Eric W. and {Konidaris}, Nicholas P. and {Herbst}, Tom and {S{\'a}nchez-Gallego}, Jos{\'e} and {Kollmeier}, Juna A. and {de Almeida}, Florence and {Barrera-Ballesteros}, Jorge K. and {Bizyaev}, Dmitry and {Brownstein}, Joel R. and {i Saguer}, Mar Canal and {Cherinka}, Brian and {Cioni}, Maria-Rosa L. and {Congiu}, Enrico and {Cosens}, Maren and {Dias}, Bruno and {Donor}, John and {Egorov}, Oleg and {Egorova}, Evgeniia and {Froning}, Cynthia S. and {Garc{\'\i}a}, Pablo and {Glover}, Simon C.~O. and {Greve}, Hannah and {H{\"a}berle}, Maximilian and {Hoy}, Kevin and {Ibarra}, Hector and {Li}, Jing and {Klessen}, Ralf S. and {Krishnarao}, Dhanesh and {Kumari}, Nimisha and {Long}, Knox S. and {M{\'e}ndez-Delgado}, Jos{\'e} Eduardo and {Popa}, Silvia Anastasia and {Ramirez}, Solange and {Rix}, Hans-Walter and {S{\'a}nchez}, Aurora Mata and {Sankrit}, Ravi and {Sattler}, Natascha and {Sayres}, Conor and {Singh}, Amrita and {Stringfellow}, Guy and {Wachter}, Stefanie and {Watkins}, Elizabeth Jayne and {Wong}, Tony and {Wofford}, Aida},
  title         = {{The SDSS-V Local Volume Mapper (LVM): Scientific Motivation and Project Overview}},
  journal       = {\aj},
  year          = {2024},
  volume        = {168},
  number        = {5},
  pages         = {198},
  month         = nov,
  adsurl        = {https://ui.adsabs.harvard.edu/abs/2024AJ....168..198D},
  archiveprefix = {arXiv},
  doi           = {10.3847/1538-3881/ad6de9},
  eid           = {198},
  eprint        = {2405.01637},
  primaryclass  = {astro-ph.GA},
}

@Article{Bica88,
  author   = {{Bica}, E.},
  title    = {{Population synthesis in galactic nuclei using a library of star clusters.}},
  journal  = {\aap},
  year     = {1988},
  volume   = {195},
  pages    = {76-92},
  month    = apr,
  adsurl   = {https://ui.adsabs.harvard.edu/abs/1988A&A...195...76B},
}

@Article{Meszaros24,
  author        = {{M{\'e}sz{\'a}ros}, Szabolcs and {Bohlin}, Ralph and {Allende Prieto}, Carlos and {Cseh}, Borb{\'a}la and {Kov{\'a}cs}, J{\'o}zsef and {Fleming}, Scott W. and {Dencs}, Zolt{\'a}n and {Deustua}, Susana and {Gordon}, Karl D. and {Hubeny}, Ivan and {Mez{\H{o}}}, Gy{\"o}rgy and {Truszek}, M{\'a}rton},
  title         = {{The updated BOSZ synthetic stellar spectral library}},
  journal       = {\aap},
  year          = {2024},
  volume        = {688},
  pages         = {A197},
  month         = aug,
  adsurl        = {https://ui.adsabs.harvard.edu/abs/2024A&A...688A.197M},
  archiveprefix = {arXiv},
  doi           = {10.1051/0004-6361/202449306},
  eid           = {A197},
  eprint        = {2407.10872},
  primaryclass  = {astro-ph.SR},
}

@Article{Hidalgo18,
  author        = {{Hidalgo}, Sebastian L. and {Pietrinferni}, Adriano and {Cassisi}, Santi and {Salaris}, Maurizio and {Mucciarelli}, Alessio and {Savino}, Alessandro and {Aparicio}, Antonio and {Silva Aguirre}, Victor and {Verma}, Kuldeep},
  title         = {{The Updated BaSTI Stellar Evolution Models and Isochrones. I. Solar-scaled Calculations}},
  journal       = {\apj},
  year          = {2018},
  volume        = {856},
  number        = {2},
  pages         = {125},
  month         = apr,
  adsurl        = {https://ui.adsabs.harvard.edu/abs/2018ApJ...856..125H},
  archiveprefix = {arXiv},
  doi           = {10.3847/1538-4357/aab158},
  eid           = {125},
  eprint        = {1802.07319},
  primaryclass  = {astro-ph.GA},
}

@InProceedings{Guido05,
  author        = {{De Marchi}, Guido and {Paresce}, Francesco and {Portegies Zwart}, Simon},
  title         = {{The Stellar IMF of Galactic Clusters and Its Evolution}},
  booktitle     = {The Initial Mass Function 50 Years Later},
  year          = {2005},
  editor        = {{Corbelli}, E. and {Palla}, F. and {Zinnecker}, H.},
  volume        = {327},
  series        = {Astrophysics and Space Science Library},
  pages         = {77},
  month         = jan,
  adsurl        = {https://ui.adsabs.harvard.edu/abs/2005ASSL..327...77D},
  archiveprefix = {arXiv},
  doi           = {10.1007/978-1-4020-3407-7_11},
  eprint        = {astro-ph/0409601},
  primaryclass  = {astro-ph},
}

@Article{Pietrinferni21,
  author        = {{Pietrinferni}, Adriano and {Hidalgo}, Sebastian and {Cassisi}, Santi and {Salaris}, Maurizio and {Savino}, Alessandro and {Mucciarelli}, Alessio and {Verma}, Kuldeep and {Silva Aguirre}, Victor and {Aparicio}, Antonio and {Ferguson}, Jason W.},
  title         = {{Updated BaSTI Stellar Evolution Models and Isochrones. II. {\ensuremath{\alpha}}-enhanced Calculations}},
  journal       = {\apj},
  year          = {2021},
  volume        = {908},
  number        = {1},
  pages         = {102},
  month         = feb,
  adsurl        = {https://ui.adsabs.harvard.edu/abs/2021ApJ...908..102P},
  archiveprefix = {arXiv},
  doi           = {10.3847/1538-4357/abd4d5},
  eid           = {102},
  eprint        = {2012.10085},
  primaryclass  = {astro-ph.SR},
}

@Article{Pietrinferni24,
  author        = {{Pietrinferni}, Adriano and {Salaris}, Maurizio and {Cassisi}, Santi and {Savino}, Alessandro and {Mucciarelli}, Alessio and {Hyder}, David and {Hidalgo}, Sebastian},
  title         = {{The updated BaSTI stellar evolution models and isochrones - IV. {\ensuremath{\alpha}}-Depleted calculations}},
  journal       = {\mnras},
  year          = {2024},
  volume        = {527},
  number        = {2},
  pages         = {2065-2070},
  month         = jan,
  adsurl        = {https://ui.adsabs.harvard.edu/abs/2024MNRAS.527.2065P},
  archiveprefix = {arXiv},
  doi           = {10.1093/mnras/stad3267},
  eprint        = {2311.05985},
  primaryclass  = {astro-ph.SR},
}

@Article{Worthey11,
  author        = {{Worthey}, Guy and {Lee}, Hyun-chul},
  title         = {{An Empirical UBV RI JHK Color-Temperature Calibration for Stars}},
  journal       = {\apjs},
  year          = {2011},
  volume        = {193},
  number        = {1},
  pages         = {1},
  month         = mar,
  adsurl        = {https://ui.adsabs.harvard.edu/abs/2011ApJS..193....1W},
  archiveprefix = {arXiv},
  doi           = {10.1088/0067-0049/193/1/1},
  eid           = {1},
  eprint        = {astro-ph/0604590},
  primaryclass  = {astro-ph},
}

@Article{Patricia24,
  author        = {{Iglesias-Navarro}, Patricia and {Huertas-Company}, Marc and {Mart{\'\i}n-Navarro}, Ignacio and {Knapen}, Johan H. and {Pernet}, Emilie},
  title         = {{Deriving the star formation histories of galaxies from spectra with simulation-based inference}},
  journal       = {\aap},
  year          = {2024},
  volume        = {689},
  pages         = {A58},
  month         = sep,
  adsurl        = {https://ui.adsabs.harvard.edu/abs/2024A&A...689A..58I},
  archiveprefix = {arXiv},
  doi           = {10.1051/0004-6361/202449909},
  eid           = {A58},
  eprint        = {2406.18661},
  primaryclass  = {astro-ph.GA},
}

@Article{Hahn22,
  author        = {{Hahn}, ChangHoon and {Melchior}, Peter},
  title         = {{Accelerated Bayesian SED Modeling Using Amortized Neural Posterior Estimation}},
  journal       = {\apj},
  year          = {2022},
  volume        = {938},
  number        = {1},
  pages         = {11},
  month         = oct,
  adsurl        = {https://ui.adsabs.harvard.edu/abs/2022ApJ...938...11H},
  archiveprefix = {arXiv},
  doi           = {10.3847/1538-4357/ac7b84},
  eid           = {11},
  eprint        = {2203.07391},
  primaryclass  = {astro-ph.GA},
}

@Article{Alsing20,
  author        = {{Alsing}, Justin and {Peiris}, Hiranya and {Leja}, Joel and {Hahn}, ChangHoon and {Tojeiro}, Rita and {Mortlock}, Daniel and {Leistedt}, Boris and {Johnson}, Benjamin D. and {Conroy}, Charlie},
  title         = {{SPECULATOR: Emulating Stellar Population Synthesis for Fast and Accurate Galaxy Spectra and Photometry}},
  journal       = {\apjs},
  year          = {2020},
  volume        = {249},
  number        = {1},
  pages         = {5},
  month         = jul,
  adsurl        = {https://ui.adsabs.harvard.edu/abs/2020ApJS..249....5A},
  archiveprefix = {arXiv},
  doi           = {10.3847/1538-4365/ab917f},
  eid           = {5},
  eprint        = {1911.11778},
  primaryclass  = {astro-ph.IM},
}

@Article{Robotham25,
  author        = {{Robotham}, A.~S.~G. and {Bellstedt}, S.},
  title         = {{ProGeny I: a new simple stellar population spectra generator and impact of isochrones / stellar spectra / initial mass functions}},
  journal       = {RAS Techniques and Instruments},
  year          = {2025},
  volume        = {4},
  pages         = {rzaf019},
  month         = jan,
  adsurl        = {https://ui.adsabs.harvard.edu/abs/2025RASTI...4...19R},
  archiveprefix = {arXiv},
  doi           = {10.1093/rasti/rzaf019},
  eid           = {rzaf019},
  eprint        = {2410.17697},
  primaryclass  = {astro-ph.GA},
}

@Article{Faber85,
  author   = {{Faber}, S.~M. and {Friel}, E.~D. and {Burstein}, D. and {Gaskell}, C.~M.},
  title    = {{Old stellar populations. II. an analysis of K-giant spectra.}},
  journal  = {\apjs},
  year     = {1985},
  volume   = {57},
  pages    = {711-741},
  month    = apr,
  adsurl   = {https://ui.adsabs.harvard.edu/abs/1985ApJS...57..711F},
  doi      = {10.1086/191024},
}

@InProceedings{jax,
  author      = {Frostig, Roy and Johnson, Matthew James and Leary, Chris},
  title       = {{Compiling Machine Learning Programs via High-Level Tracing}},
  booktitle   = {{SysML Conference 2018}},
  year        = {2019},
  address     = {Stanford, United States},
  month       = Mar,
  hal_id      = {hal-05188750},
  hal_version = {v1},
  url         = {https://hal.science/hal-05188750},
}

@Software{jax2018github,
  author  = {James Bradbury and Roy Frostig and Peter Hawkins and Matthew James Johnson and Chris Leary and Dougal Maclaurin and George Necula and Adam Paszke and Jake Vander{P}las and Skye Wanderman-{M}ilne and Qiao Zhang},
  title   = {{JAX}: composable transformations of {P}ython+{N}um{P}y programs},
  year    = {2018},
  url     = {http://github.com/jax-ml/jax},
  version = {0.3.13},
}

@Article{MN24b,
  author        = {{Mart{\'\i}n-Navarro}, I. and {Vazdekis}, A.},
  title         = {{Counting stars from the integrated spectra of galaxies}},
  journal       = {\aap},
  year          = {2024},
  volume        = {691},
  pages         = {L10},
  month         = nov,
  adsurl        = {https://ui.adsabs.harvard.edu/abs/2024A&A...691L..10M},
  archiveprefix = {arXiv},
  doi           = {10.1051/0004-6361/202451604},
  eid           = {L10},
  eprint        = {2410.11961},
  primaryclass  = {astro-ph.GA},
}

@InProceedings{Bruzual02,
  author        = {{Bruzual A.}, Gustavo},
  title         = {{Stellar Populations in Star Clusters: The R{\^o}le Played by Stochastic Effects}},
  booktitle     = {Extragalactic Star Clusters},
  year          = {2002},
  editor        = {{Geisler}, Douglas Paul and {Grebel}, Eva K. and {Minniti}, Dante},
  volume        = {207},
  series        = {IAU Symposium},
  pages         = {616},
  month         = jan,
  adsurl        = {https://ui.adsabs.harvard.edu/abs/2002IAUS..207..616B},
  archiveprefix = {arXiv},
  doi           = {10.48550/arXiv.astro-ph/0110245},
  eprint        = {astro-ph/0110245},
  primaryclass  = {astro-ph},
}

@Article{Miguel06,
  author        = {{Cervi{\~n}o}, M. and {Luridiana}, V.},
  title         = {{Confidence limits of evolutionary synthesis models. IV. Moving forward to a probabilistic formulation}},
  journal       = {\aap},
  year          = {2006},
  volume        = {451},
  number        = {2},
  pages         = {475-498},
  month         = may,
  adsurl        = {https://ui.adsabs.harvard.edu/abs/2006A&A...451..475C},
  archiveprefix = {arXiv},
  doi           = {10.1051/0004-6361:20053283},
  eprint        = {astro-ph/0504483},
  primaryclass  = {astro-ph},
}

@Article{Miguel02,
  author        = {{Cervi{\~n}o}, M. and {Valls-Gabaud}, D. and {Luridiana}, V. and {Mas-Hesse}, J.~M.},
  title         = {{Confidence levels of evolutionary synthesis models. II. On sampling and Poissonian fluctuations}},
  journal       = {\aap},
  year          = {2002},
  volume        = {381},
  pages         = {51-64},
  month         = jan,
  adsurl        = {https://ui.adsabs.harvard.edu/abs/2002A&A...381...51C},
  archiveprefix = {arXiv},
  doi           = {10.1051/0004-6361:20011266},
  eprint        = {astro-ph/0109435},
  primaryclass  = {astro-ph},
}

@Article{Renzini98,
  author        = {{Renzini}, Alvio},
  title         = {{The Stellar Populations of Pixels and Frames}},
  journal       = {\aj},
  year          = {1998},
  volume        = {115},
  number        = {6},
  pages         = {2459-2465},
  month         = jun,
  adsurl        = {https://ui.adsabs.harvard.edu/abs/1998AJ....115.2459R},
  archiveprefix = {arXiv},
  doi           = {10.1086/300356},
  eprint        = {astro-ph/9802186},
  primaryclass  = {astro-ph},
}

@Article{Conroy16,
  author        = {{Conroy}, Charlie and {van Dokkum}, Pieter G.},
  title         = {{Pixel Color Magnitude Diagrams for Semi-resolved Stellar Populations: The Star Formation History of Regions within the Disk and Bulge of M31}},
  journal       = {\apj},
  year          = {2016},
  volume        = {827},
  number        = {1},
  pages         = {9},
  month         = aug,
  adsurl        = {https://ui.adsabs.harvard.edu/abs/2016ApJ...827....9C},
  archiveprefix = {arXiv},
  doi           = {10.3847/0004-637X/827/1/9},
  eid           = {9},
  eprint        = {1602.05580},
  primaryclass  = {astro-ph.GA},
}

@Article{Krumholz15,
  author        = {{Krumholz}, Mark R. and {Fumagalli}, Michele and {da Silva}, Robert L. and {Rendahl}, Theodore and {Parra}, Jonathan},
  title         = {{SLUG - stochastically lighting up galaxies - III. A suite of tools for simulated photometry, spectroscopy, and Bayesian inference with stochastic stellar populations}},
  journal       = {\mnras},
  year          = {2015},
  volume        = {452},
  number        = {2},
  pages         = {1447-1467},
  month         = sep,
  adsurl        = {https://ui.adsabs.harvard.edu/abs/2015MNRAS.452.1447K},
  archiveprefix = {arXiv},
  doi           = {10.1093/mnras/stv1374},
  eprint        = {1502.05408},
  primaryclass  = {astro-ph.GA},
}

@Article{dS12,
  author        = {{da Silva}, Robert L. and {Fumagalli}, Michele and {Krumholz}, Mark},
  title         = {{SLUG{\textemdash}Stochastically Lighting Up Galaxies. I. Methods and Validating Tests}},
  journal       = {\apj},
  year          = {2012},
  volume        = {745},
  number        = {2},
  pages         = {145},
  month         = feb,
  adsurl        = {https://ui.adsabs.harvard.edu/abs/2012ApJ...745..145D},
  archiveprefix = {arXiv},
  doi           = {10.1088/0004-637X/745/2/145},
  eid           = {145},
  eprint        = {1106.3072},
  primaryclass  = {astro-ph.IM},
}

@Article{Orozco-Duarte,
  author        = {{Orozco-Duarte}, Rogelio and {Wofford}, Aida and {Vidal-Garc{\'\i}a}, Alba and {Bruzual}, Gustavo and {Charlot}, Stephane and {Krumholz}, Mark R. and {Hannon}, Stephen and {Lee}, Janice and {Wofford}, Timothy and {Fumagalli}, Michele and {Dale}, Daniel and {Messa}, Matteo and {Grebel}, Eva K. and {Smith}, Linda and {Grasha}, Kathryn and {Cook}, David},
  title         = {{Synthetic photometry of OB star clusters with stochastically sampled IMFs: analysis of models and HST observations}},
  journal       = {\mnras},
  year          = {2022},
  volume        = {509},
  number        = {1},
  pages         = {522-549},
  month         = jan,
  adsurl        = {https://ui.adsabs.harvard.edu/abs/2022MNRAS.509..522O},
  archiveprefix = {arXiv},
  doi           = {10.1093/mnras/stab2988},
  eprint        = {2110.05595},
  primaryclass  = {astro-ph.GA},
}

@Article{Eldridge12,
  author        = {{Eldridge}, John J.},
  title         = {{Stochasticity, a variable stellar upper mass limit, binaries and star formation rate indicators}},
  journal       = {\mnras},
  year          = {2012},
  volume        = {422},
  number        = {1},
  pages         = {794-803},
  month         = may,
  adsurl        = {https://ui.adsabs.harvard.edu/abs/2012MNRAS.422..794E},
  archiveprefix = {arXiv},
  doi           = {10.1111/j.1365-2966.2012.20662.x},
  eprint        = {1106.4311},
  primaryclass  = {astro-ph.SR},
}

@Article{Stanway23,
  author        = {{Stanway}, Elizabeth R. and {Eldridge}, J.~J.},
  title         = {{Exploring the impact of IMF and binary parameter stochasticity with a binary population synthesis code}},
  journal       = {\mnras},
  year          = {2023},
  volume        = {522},
  number        = {3},
  pages         = {4430-4443},
  month         = jul,
  adsurl        = {https://ui.adsabs.harvard.edu/abs/2023MNRAS.522.4430S},
  archiveprefix = {arXiv},
  doi           = {10.1093/mnras/stad1185},
  eprint        = {2304.09549},
  primaryclass  = {astro-ph.GA},
}

@Article{Branco24,
  author        = {{Branco}, Vinicius and {Coelho}, Paula R.~T. and {Lan{\c{c}}on}, Ariane and {Martins}, Lucimara P. and {Prugniel}, Philippe},
  title         = {{Synthetic stellar spectra for studying multiple populations in globular clusters. Extended grid, and the effects on the integrated light}},
  journal       = {\aap},
  year          = {2024},
  volume        = {687},
  pages         = {A142},
  month         = jul,
  adsurl        = {https://ui.adsabs.harvard.edu/abs/2024A&A...687A.142B},
  archiveprefix = {arXiv},
  doi           = {10.1051/0004-6361/202348992},
  eid           = {A142},
  eprint        = {2404.15468},
  primaryclass  = {astro-ph.SR},
}

@Article{Beerman12,
  author        = {{Beerman}, Lori C. and {Johnson}, L. Clifton and {Fouesneau}, Morgan and {Dalcanton}, Julianne J. and {Weisz}, Daniel R. and {Seth}, Anil C. and {Williams}, Ben F. and {Bell}, Eric F. and {Bianchi}, Luciana C. and {Caldwell}, Nelson and {Dolphin}, Andrew E. and {Gouliermis}, Dimitrios A. and {Kalirai}, Jason S. and {Larsen}, S{\o}ren S. and {Melbourne}, Jason L. and {Rix}, Hans-Walter and {Skillman}, Evan D.},
  title         = {{The Panchromatic Hubble Andromeda Treasury. III. Measuring Ages and Masses of Partially Resolved Stellar Clusters}},
  journal       = {\apj},
  year          = {2012},
  volume        = {760},
  number        = {2},
  pages         = {104},
  month         = dec,
  adsurl        = {https://ui.adsabs.harvard.edu/abs/2012ApJ...760..104B},
  archiveprefix = {arXiv},
  doi           = {10.1088/0004-637X/760/2/104},
  eid           = {104},
  eprint        = {1209.5395},
  primaryclass  = {astro-ph.CO},
}

@Article{Fouesneau10,
  author        = {{Fouesneau}, M. and {Lan{\c{c}}on}, A.},
  title         = {{Accounting for stochastic fluctuations when analysing the integrated light of star clusters. I. First systematics}},
  journal       = {\aap},
  year          = {2010},
  volume        = {521},
  pages         = {A22},
  month         = oct,
  adsurl        = {https://ui.adsabs.harvard.edu/abs/2010A&A...521A..22F},
  archiveprefix = {arXiv},
  doi           = {10.1051/0004-6361/201014084},
  eid           = {A22},
  eprint        = {1003.2334},
  primaryclass  = {astro-ph.SR},
}

@Article{Pablo21,
  author        = {{Rodr{\'\i}guez-Beltr{\'a}n}, P. and {Vazdekis}, A. and {Cervi{\~n}o}, M. and {Beasley}, M.~A.},
  title         = {{Surface brightness fluctuations to constrain secondary stellar populations: revealing very low-metallicity stars in massive galaxies}},
  journal       = {\mnras},
  year          = {2021},
  volume        = {507},
  number        = {2},
  pages         = {3005-3029},
  month         = oct,
  adsurl        = {https://ui.adsabs.harvard.edu/abs/2021MNRAS.507.3005R},
  archiveprefix = {arXiv},
  doi           = {10.1093/mnras/stab2330},
  eprint        = {2107.08722},
  primaryclass  = {astro-ph.GA},
}

@Article{Tonry88,
  author   = {{Tonry}, John and {Schneider}, Donald P.},
  title    = {{A New Technique for Measuring Extragalactic Distances}},
  journal  = {\aj},
  year     = {1988},
  volume   = {96},
  pages    = {807},
  month    = sep,
  adsurl   = {https://ui.adsabs.harvard.edu/abs/1988AJ.....96..807T},
  doi      = {10.1086/114847},
}

@Article{Buzzoni93,
  author  = {{Buzzoni}, A.},
  title   = {{Statistical properties of stellar populations and surface-brightness fluctuations in galaxies.}},
  journal = {\aap},
  year    = {1993},
  volume  = {275},
  pages   = {433-450},
  month   = aug,
  adsurl  = {https://ui.adsabs.harvard.edu/abs/1993A&A...275..433B},
}

@Article{Tonry01,
  author        = {{Tonry}, John L. and {Dressler}, Alan and {Blakeslee}, John P. and {Ajhar}, Edward A. and {Fletcher}, Andr{\'e} B. and {Luppino}, Gerard A. and {Metzger}, Mark R. and {Moore}, Christopher B.},
  title         = {{The SBF Survey of Galaxy Distances. IV. SBF Magnitudes, Colors, and Distances}},
  journal       = {\apj},
  year          = {2001},
  volume        = {546},
  number        = {2},
  pages         = {681-693},
  month         = jan,
  adsurl        = {https://ui.adsabs.harvard.edu/abs/2001ApJ...546..681T},
  archiveprefix = {arXiv},
  doi           = {10.1086/318301},
  eprint        = {astro-ph/0011223},
  primaryclass  = {astro-ph},
}

@Article{Fouesneau12,
  author        = {{Fouesneau}, Morgan and {Lan{\c{c}}on}, Ariane and {Chandar}, Rupali and {Whitmore}, Bradley C.},
  title         = {{Analyzing Star Cluster Populations with Stochastic Models: The Hubble Space Telescope/Wide Field Camera 3 Sample of Clusters in M83}},
  journal       = {\apj},
  year          = {2012},
  volume        = {750},
  number        = {1},
  pages         = {60},
  month         = may,
  adsurl        = {https://ui.adsabs.harvard.edu/abs/2012ApJ...750...60F},
  archiveprefix = {arXiv},
  doi           = {10.1088/0004-637X/750/1/60},
  eid           = {60},
  eprint        = {1202.3135},
  primaryclass  = {astro-ph.CO},
}

@Article{Abraham14,
  author        = {{Abraham}, Roberto G. and {van Dokkum}, Pieter G.},
  title         = {{Ultra-Low Surface Brightness Imaging with the Dragonfly Telephoto Array}},
  journal       = {\pasp},
  year          = {2014},
  volume        = {126},
  number        = {935},
  pages         = {55},
  month         = jan,
  adsurl        = {https://ui.adsabs.harvard.edu/abs/2014PASP..126...55A},
  archiveprefix = {arXiv},
  doi           = {10.1086/674875},
  eprint        = {1401.5473},
  primaryclass  = {astro-ph.IM},
}

@Article{Trujillo16,
  author        = {{Trujillo}, Ignacio and {Fliri}, J{\"u}ergen},
  title         = {{Beyond 31 mag arcsec$^{-2}$: The Frontier of Low Surface Brightness Imaging with the Largest Optical Telescopes}},
  journal       = {\apj},
  year          = {2016},
  volume        = {823},
  number        = {2},
  pages         = {123},
  month         = jun,
  adsurl        = {https://ui.adsabs.harvard.edu/abs/2016ApJ...823..123T},
  archiveprefix = {arXiv},
  doi           = {10.3847/0004-637X/823/2/123},
  eid           = {123},
  eprint        = {1510.04696},
  primaryclass  = {astro-ph.GA},
}

@Article{Trujillo21,
  author        = {{Trujillo}, Ignacio and {D'Onofrio}, Mauro and {Zaritsky}, Dennis and {Madrigal-Aguado}, Alberto and {Chamba}, Nushkia and {Golini}, Giulia and {Akhlaghi}, Mohammad and {Sharbaf}, Zahra and {Infante-Sainz}, Ra{\'u}l and {Rom{\'a}n}, Javier and {Morales-Socorro}, Carlos and {Sand}, David J. and {Martin}, Garreth},
  title         = {{Introducing the LBT Imaging of Galactic Halos and Tidal Structures (LIGHTS) survey. A preview of the low surface brightness Universe to be unveiled by LSST}},
  journal       = {\aap},
  year          = {2021},
  volume        = {654},
  pages         = {A40},
  month         = oct,
  adsurl        = {https://ui.adsabs.harvard.edu/abs/2021A&A...654A..40T},
  archiveprefix = {arXiv},
  doi           = {10.1051/0004-6361/202141603},
  eid           = {A40},
  eprint        = {2109.07478},
  primaryclass  = {astro-ph.GA},
}

@Article{Mihos17,
  author        = {{Mihos}, J. Christopher and {Harding}, Paul and {Feldmeier}, John J. and {Rudick}, Craig and {Janowiecki}, Steven and {Morrison}, Heather and {Slater}, Colin and {Watkins}, Aaron},
  title         = {{The Burrell Schmidt Deep Virgo Survey: Tidal Debris, Galaxy Halos, and Diffuse Intracluster Light in the Virgo Cluster}},
  journal       = {\apj},
  year          = {2017},
  volume        = {834},
  number        = {1},
  pages         = {16},
  month         = jan,
  adsurl        = {https://ui.adsabs.harvard.edu/abs/2017ApJ...834...16M},
  archiveprefix = {arXiv},
  doi           = {10.3847/1538-4357/834/1/16},
  eid           = {16},
  eprint        = {1611.04435},
  primaryclass  = {astro-ph.GA},
}

@Article{vdk15,
  author        = {{van Dokkum}, Pieter G. and {Abraham}, Roberto and {Merritt}, Allison and {Zhang}, Jielai and {Geha}, Marla and {Conroy}, Charlie},
  title         = {{Forty-seven Milky Way-sized, Extremely Diffuse Galaxies in the Coma Cluster}},
  journal       = {\apjl},
  year          = {2015},
  volume        = {798},
  number        = {2},
  pages         = {L45},
  month         = jan,
  adsurl        = {https://ui.adsabs.harvard.edu/abs/2015ApJ...798L..45V},
  archiveprefix = {arXiv},
  doi           = {10.1088/2041-8205/798/2/L45},
  eid           = {L45},
  eprint        = {1410.8141},
  primaryclass  = {astro-ph.GA},
}

@Article{Euclid,
  author        = {{Euclid Collaboration} and {Urbano}, M. and {Duc}, P.-A. and {Poulain}, M. and {Nucita}, A.~A. and {Venhola}, A. and {Marchal}, O. and {K{\"u}mmel}, M. and {Kong}, H. and {Soldano}, F. and {Romelli}, E. and {Walmsley}, M. and {Saifollahi}, T. and {Voggel}, K. and {Lan{\c{c}}on}, A. and {Marleau}, F.~R. and {Sola}, E. and {Hunt}, L.~K. and {Junais}, J. and {Carollo}, D. and {Sanchez-Alarcon}, P.~M. and {Baes}, M. and {Buitrago}, F. and {Cantiello}, M. and {Cuillandre}, J.-C. and {Dom{\'\i}nguez S{\'a}nchez}, H. and {Ferr{\'e}-Mateu}, A. and {Franco}, A. and {Gracia-Carpio}, J. and {Habas}, R. and {Hilker}, M. and {Iodice}, E. and {Knapen}, J.~H. and {Le}, M.~N. and {Mart{\'\i}nez-Delgado}, D. and {M{\"u}ller}, O. and {De Paolis}, F. and {Papaderos}, P. and {Ragusa}, R. and {Rom{\'a}n}, J. and {Saremi}, E. and {Testa}, V. and {Altieri}, B. and {Amendola}, L. and {Andreon}, S. and {Auricchio}, N. and {Baccigalupi}, C. and {Baldi}, M. and {Bardelli}, S. and {Battaglia}, P. and {Biviano}, A. and {Branchini}, E. and {Brescia}, M. and {Camera}, S. and {Ca{\~n}as-Herrera}, G. and {Capobianco}, V. and {Carbone}, C. and {Carretero}, J. and {Casas}, S. and {Castellano}, M. and {Castignani}, G. and {Cavuoti}, S. and {Cimatti}, A. and {Colodro-Conde}, C. and {Congedo}, G. and {Conselice}, C.~J. and {Conversi}, L. and {Copin}, Y. and {Courbin}, F. and {Courtois}, H.~M. and {Cropper}, M. and {Da Silva}, A. and {Degaudenzi}, H. and {De Lucia}, G. and {Dole}, H. and {Dubath}, F. and {Duncan}, C.~A.~J. and {Dupac}, X. and {Dusini}, S. and {Escoffier}, S. and {Farina}, M. and {Farinelli}, R. and {Ferriol}, S. and {Finelli}, F. and {Frailis}, M. and {Franceschi}, E. and {Fumana}, M. and {Galeotta}, S. and {George}, K. and {Gillis}, B. and {Giocoli}, C. and {Grazian}, A. and {Grupp}, F. and {Guzzo}, L. and {Haugan}, S.~V.~H. and {Holmes}, W. and {Hook}, I.~M. and {Hormuth}, F. and {Hornstrup}, A. and {Jahnke}, K. and {Jhabvala}, M. and {Joachimi}, B. and {Keih{\"a}nen}, E. and {Kermiche}, S. and {Kiessling}, A. and {Kubik}, B. and {Kunz}, M. and {Kurki-Suonio}, H. and {Laureijs}, R. and {Le Brun}, A.~M.~C. and {Ligori}, S. and {Lilje}, P.~B. and {Lindholm}, V. and {Lloro}, I. and {Mainetti}, G. and {Maino}, D. and {Maiorano}, E. and {Mansutti}, O. and {Marggraf}, O. and {Martinelli}, M. and {Martinet}, N. and {Marulli}, F. and {Massey}, R.~J. and {Medinaceli}, E. and {Mei}, S. and {Mellier}, Y. and {Meneghetti}, M. and {Merlin}, E. and {Meylan}, G. and {Mora}, A. and {Moresco}, M. and {Moscardini}, L. and {Nakajima}, R. and {Neissner}, C. and {Niemi}, S.-M. and {Padilla}, C. and {Paltani}, S. and {Pasian}, F. and {Pedersen}, K. and {Pettorino}, V. and {Pires}, S. and {Polenta}, G. and {Poncet}, M. and {Popa}, L.~A. and {Pozzetti}, L. and {Raison}, F. and {Rebolo}, R. and {Renzi}, A. and {Rhodes}, J. and {Riccio}, G. and {Roncarelli}, M. and {Saglia}, R. and {Sakr}, Z. and {Sapone}, D. and {Sartoris}, B. and {Schneider}, P. and {Schrabback}, T. and {Secroun}, A. and {Seidel}, G. and {Serrano}, S. and {Simon}, P. and {Sirignano}, C. and {Sirri}, G. and {Stanco}, L. and {Starck}, J.-L. and {Steinwagner}, J. and {Tallada-Cresp{\'\i}}, P. and {Taylor}, A.~N. and {Teplitz}, H.~I. and {Tereno}, I. and {Tessore}, N. and {Toft}, S. and {Toledo-Moreo}, R. and {Torradeflot}, F. and {Tutusaus}, I. and {Valenziano}, L. and {Valiviita}, J. and {Vassallo}, T. and {Verdoes Kleijn}, G. and {Veropalumbo}, A. and {Wang}, Y. and {Weller}, J. and {Zamorani}, G. and {Zinchenko}, I.~A. and {Zucca}, E. and {Ballardini}, M. and {Bolzonella}, M. and {Bozzo}, E. and {Burigana}, C. and {Cabanac}, R. and {Cappi}, A. and {Di Ferdinando}, D. and {Escartin Vigo}, J.~A. and {Gabarra}, L. and {Huertas-Company}, M. and {Mart{\'\i}n-Fleitas}, J. and {Matthew}, S. and {Mauri}, N. and {Metcalf}, R.~B. and {Pezzotta}, A.},
  title         = {{Euclid preparation: LXXIX. Using mock low surface brightness dwarf galaxies to probe Euclid Wide Survey detection capabilities}},
  journal       = {\aap},
  year          = {2026},
  volume        = {707},
  pages         = {A229},
  month         = mar,
  adsurl        = {https://ui.adsabs.harvard.edu/abs/2026A&A...707A.229E},
  archiveprefix = {arXiv},
  doi           = {10.1051/0004-6361/202557270},
  eid           = {A229},
  eprint        = {2509.13163},
  primaryclass  = {astro-ph.GA},
}

@Article{Euclid0,
  author        = {{Euclid Collaboration} and {Mellier}, Y. and {Abdurro'uf} and {Acevedo Barroso}, J.~A. and {Ach{\'u}carro}, A. and {Adamek}, J. and {Adam}, R. and {Addison}, G.~E. and {Aghanim}, N. and {Aguena}, M. and {Ajani}, V. and {Akrami}, Y. and {Al-Bahlawan}, A. and {Alavi}, A. and {Albuquerque}, I.~S. and {Alestas}, G. and {Alguero}, G. and {Allaoui}, A. and {Allen}, S.~W. and {Allevato}, V. and {Alonso-Tetilla}, A.~V. and {Altieri}, B. and {Alvarez-Candal}, A. and {Alvi}, S. and {Amara}, A. and {Amendola}, L. and {Amiaux}, J. and {Andika}, I.~T. and {Andreon}, S. and {Andrews}, A. and {Angora}, G. and {Angulo}, R.~E. and {Annibali}, F. and {Anselmi}, A. and {Anselmi}, S. and {Arcari}, S. and {Archidiacono}, M. and {Aric{\`o}}, G. and {Arnaud}, M. and {Arnouts}, S. and {Asgari}, M. and {Asorey}, J. and {Atayde}, L. and {Atek}, H. and {Atrio-Barandela}, F. and {Aubert}, M. and {Aubourg}, E. and {Auphan}, T. and {Auricchio}, N. and {Aussel}, B. and {Aussel}, H. and {Avelino}, P.~P. and {Avgoustidis}, A. and {Avila}, S. and {Awan}, S. and {Azzollini}, R. and {Baccigalupi}, C. and {Bachelet}, E. and {Bacon}, D. and {Baes}, M. and {Bagley}, M.~B. and {Bahr-Kalus}, B. and {Balaguera-Antolinez}, A. and {Balbinot}, E. and {Balcells}, M. and {Baldi}, M. and {Baldry}, I. and {Balestra}, A. and {Ballardini}, M. and {Ballester}, O. and {Balogh}, M. and {Ba{\~n}ados}, E. and {Barbier}, R. and {Bardelli}, S. and {Baron}, M. and {Barreiro}, T. and {Barrena}, R. and {Barriere}, J.-C. and {Barros}, B.~J. and {Barthelemy}, A. and {Bartolo}, N. and {Basset}, A. and {Battaglia}, P. and {Battisti}, A.~J. and {Baugh}, C.~M. and {Baumont}, L. and {Bazzanini}, L. and {Beaulieu}, J.-P. and {Beckmann}, V. and {Belikov}, A.~N. and {Bel}, J. and {Bellagamba}, F. and {Bella}, M. and {Bellini}, E. and {Benabed}, K. and {Bender}, R. and {Benevento}, G. and {Bennett}, C.~L. and {Benson}, K. and {Bergamini}, P. and {Bermejo-Climent}, J.~R. and {Bernardeau}, F. and {Bertacca}, D. and {Berthe}, M. and {Berthier}, J. and {Bethermin}, M. and {Beutler}, F. and {Bevillon}, C. and {Bhargava}, S. and {Bhatawdekar}, R. and {Bianchi}, D. and {Bisigello}, L. and {Biviano}, A. and {Blake}, R.~P. and {Blanchard}, A. and {Blazek}, J. and {Blot}, L. and {Bosco}, A. and {Bodendorf}, C. and {Boenke}, T. and {B{\"o}hringer}, H. and {Boldrini}, P. and {Bolzonella}, M. and {Bonchi}, A. and {Bonici}, M. and {Bonino}, D. and {Bonino}, L. and {Bonvin}, C. and {Bon}, W. and {Booth}, J.~T. and {Borgani}, S. and {Borlaff}, A.~S. and {Borsato}, E. and {Bose}, B. and {Botticella}, M.~T. and {Boucaud}, A. and {Bouche}, F. and {Boucher}, J.~S. and {Boutigny}, D. and {Bouvard}, T. and {Bouwens}, R. and {Bouy}, H. and {Bowler}, R.~A.~A. and {Bozza}, V. and {Bozzo}, E. and {Branchini}, E. and {Brando}, G. and {Brau-Nogue}, S. and {Brekke}, P. and {Bremer}, M.~N. and {Brescia}, M. and {Breton}, M.-A. and {Brinchmann}, J. and {Brinckmann}, T. and {Brockley-Blatt}, C. and {Brodwin}, M. and {Brouard}, L. and {Brown}, M.~L. and {Bruton}, S. and {Bucko}, J. and {Buddelmeijer}, H. and {Buenadicha}, G. and {Buitrago}, F. and {Burger}, P. and {Burigana}, C. and {Busillo}, V. and {Busonero}, D. and {Cabanac}, R. and {Cabayol-Garcia}, L. and {Cagliari}, M.~S. and {Caillat}, A. and {Caillat}, L. and {Calabrese}, M. and {Calabro}, A. and {Calderone}, G. and {Calura}, F. and {Camacho Quevedo}, B. and {Camera}, S. and {Campos}, L. and {Ca{\~n}as-Herrera}, G. and {Candini}, G.~P. and {Cantiello}, M. and {Capobianco}, V. and {Cappellaro}, E. and {Cappelluti}, N. and {Cappi}, A. and {Caputi}, K.~I. and {Cara}, C. and {Carbone}, C. and {Cardone}, V.~F. and {Carella}, E. and {Carlberg}, R.~G. and {Carle}, M. and {Carminati}, L. and {Caro}, F. and {Carrasco}, J.~M. and {Carretero}, J. and {Carrilho}, P. and {Carron Duque}, J. and {Carry}, B.},
  title         = {{Euclid: I. Overview of the Euclid mission}},
  journal       = {\aap},
  year          = {2025},
  volume        = {697},
  pages         = {A1},
  month         = may,
  adsurl        = {https://ui.adsabs.harvard.edu/abs/2025A&A...697A...1E},
  archiveprefix = {arXiv},
  doi           = {10.1051/0004-6361/202450810},
  eid           = {A1},
  eprint        = {2405.13491},
  primaryclass  = {astro-ph.CO},
}

@Article{Martin22,
  author        = {{Martin}, G. and {Bazkiaei}, A.~E. and {Spavone}, M. and {Iodice}, E. and {Mihos}, J.~C. and {Montes}, M. and {Benavides}, J.~A. and {Brough}, S. and {Carlin}, J.~L. and {Collins}, C.~A. and {Duc}, P.~A. and {G{\'o}mez}, F.~A. and {Galaz}, G. and {Hern{\'a}ndez-Toledo}, H.~M. and {Jackson}, R.~A. and {Kaviraj}, S. and {Knapen}, J.~H. and {Mart{\'\i}nez-Lombilla}, C. and {McGee}, S. and {O'Ryan}, D. and {Prole}, D.~J. and {Rich}, R.~M. and {Rom{\'a}n}, J. and {Shah}, E.~A. and {Starkenburg}, T.~K. and {Watkins}, A.~E. and {Zaritsky}, D. and {Pichon}, C. and {Armus}, L. and {Bianconi}, M. and {Buitrago}, F. and {Bus{\'a}}, I. and {Davis}, F. and {Demarco}, R. and {Desmons}, A. and {Garc{\'\i}a}, P. and {Graham}, A.~W. and {Holwerda}, B. and {Hon}, D.~S.-H. and {Khalid}, A. and {Klehammer}, J. and {Klutse}, D.~Y. and {Lazar}, I. and {Nair}, P. and {Noakes-Kettel}, E.~A. and {Rutkowski}, M. and {Saha}, K. and {Sahu}, N. and {Sola}, E. and {V{\'a}zquez-Mata}, J.~A. and {Vera-Casanova}, A. and {Yoon}, I.},
  title         = {{Preparing for low surface brightness science with the Vera C. Rubin Observatory: Characterization of tidal features from mock images}},
  journal       = {\mnras},
  year          = {2022},
  volume        = {513},
  number        = {1},
  pages         = {1459-1487},
  month         = jun,
  adsurl        = {https://ui.adsabs.harvard.edu/abs/2022MNRAS.513.1459M},
  archiveprefix = {arXiv},
  doi           = {10.1093/mnras/stac1003},
  eprint        = {2203.07675},
  primaryclass  = {astro-ph.GA},
}

@Article{desi,
  author        = {{Dey}, Arjun and {Schlegel}, David J. and {Lang}, Dustin and {Blum}, Robert and {Burleigh}, Kaylan and {Fan}, Xiaohui and {Findlay}, Joseph R. and {Finkbeiner}, Doug and {Herrera}, David and {Juneau}, St{\'e}phanie and {Landriau}, Martin and {Levi}, Michael and {McGreer}, Ian and {Meisner}, Aaron and {Myers}, Adam D. and {Moustakas}, John and {Nugent}, Peter and {Patej}, Anna and {Schlafly}, Edward F. and {Walker}, Alistair R. and {Valdes}, Francisco and {Weaver}, Benjamin A. and {Y{\`e}che}, Christophe and {Zou}, Hu and {Zhou}, Xu and {Abareshi}, Behzad and {Abbott}, T.~M.~C. and {Abolfathi}, Bela and {Aguilera}, C. and {Alam}, Shadab and {Allen}, Lori and {Alvarez}, A. and {Annis}, James and {Ansarinejad}, Behzad and {Aubert}, Marie and {Beechert}, Jacqueline and {Bell}, Eric F. and {BenZvi}, Segev Y. and {Beutler}, Florian and {Bielby}, Richard M. and {Bolton}, Adam S. and {Brice{\~n}o}, C{\'e}sar and {Buckley-Geer}, Elizabeth J. and {Butler}, Karen and {Calamida}, Annalisa and {Carlberg}, Raymond G. and {Carter}, Paul and {Casas}, Ricard and {Castander}, Francisco J. and {Choi}, Yumi and {Comparat}, Johan and {Cukanovaite}, Elena and {Delubac}, Timoth{\'e}e and {DeVries}, Kaitlin and {Dey}, Sharmila and {Dhungana}, Govinda and {Dickinson}, Mark and {Ding}, Zhejie and {Donaldson}, John B. and {Duan}, Yutong and {Duckworth}, Christopher J. and {Eftekharzadeh}, Sarah and {Eisenstein}, Daniel J. and {Etourneau}, Thomas and {Fagrelius}, Parker A. and {Farihi}, Jay and {Fitzpatrick}, Mike and {Font-Ribera}, Andreu and {Fulmer}, Leah and {G{\"a}nsicke}, Boris T. and {Gaztanaga}, Enrique and {George}, Koshy and {Gerdes}, David W. and {Gontcho}, Satya Gontcho A. and {Gorgoni}, Claudio and {Green}, Gregory and {Guy}, Julien and {Harmer}, Diane and {Hernandez}, M. and {Honscheid}, Klaus and {Huang}, Lijuan Wendy and {James}, David J. and {Jannuzi}, Buell T. and {Jiang}, Linhua and {Joyce}, Richard and {Karcher}, Armin and {Karkar}, Sonia and {Kehoe}, Robert and {Kneib}, Jean-Paul and {Kueter-Young}, Andrea and {Lan}, Ting-Wen and {Lauer}, Tod R. and {Le Guillou}, Laurent and {Le Van Suu}, Auguste and {Lee}, Jae Hyeon and {Lesser}, Michael and {Perreault Levasseur}, Laurence and {Li}, Ting S. and {Mann}, Justin L. and {Marshall}, Robert and {Mart{\'\i}nez-V{\'a}zquez}, C.~E. and {Martini}, Paul and {du Mas des Bourboux}, H{\'e}lion and {McManus}, Sean and {Meier}, Tobias Gabriel and {M{\'e}nard}, Brice and {Metcalfe}, Nigel and {Mu{\~n}oz-Guti{\'e}rrez}, Andrea and {Najita}, Joan and {Napier}, Kevin and {Narayan}, Gautham and {Newman}, Jeffrey A. and {Nie}, Jundan and {Nord}, Brian and {Norman}, Dara J. and {Olsen}, Knut A.~G. and {Paat}, Anthony and {Palanque-Delabrouille}, Nathalie and {Peng}, Xiyan and {Poppett}, Claire L. and {Poremba}, Megan R. and {Prakash}, Abhishek and {Rabinowitz}, David and {Raichoor}, Anand and {Rezaie}, Mehdi and {Robertson}, A.~N. and {Roe}, Natalie A. and {Ross}, Ashley J. and {Ross}, Nicholas P. and {Rudnick}, Gregory and {Safonova}, Sasha and {Saha}, Abhijit and {S{\'a}nchez}, F. Javier and {Savary}, Elodie and {Schweiker}, Heidi and {Scott}, Adam and {Seo}, Hee-Jong and {Shan}, Huanyuan and {Silva}, David R. and {Slepian}, Zachary and {Soto}, Christian and {Sprayberry}, David and {Staten}, Ryan and {Stillman}, Coley M. and {Stupak}, Robert J. and {Summers}, David L. and {Sien Tie}, Suk and {Tirado}, H. and {Vargas-Maga{\~n}a}, Mariana and {Vivas}, A. Katherina and {Wechsler}, Risa H. and {Williams}, Doug and {Yang}, Jinyi and {Yang}, Qian and {Yapici}, Tolga and {Zaritsky}, Dennis and {Zenteno}, A. and {Zhang}, Kai and {Zhang}, Tianmeng and {Zhou}, Rongpu and {Zhou}, Zhimin},
  title         = {{Overview of the DESI Legacy Imaging Surveys}},
  journal       = {\aj},
  year          = {2019},
  volume        = {157},
  number        = {5},
  pages         = {168},
  month         = may,
  adsurl        = {https://ui.adsabs.harvard.edu/abs/2019AJ....157..168D},
  archiveprefix = {arXiv},
  doi           = {10.3847/1538-3881/ab089d},
  eid           = {168},
  eprint        = {1804.08657},
  primaryclass  = {astro-ph.IM},
}

@Article{MD25,
  author        = {{Mart{\'\i}nez-Delgado}, David and {Stein}, Michael and {Sakowska}, Joanna D. and {Maurice Weigelt}, M. and {Rom{\'a}n}, Javier and {Donatiello}, Giuseppe and {Roca-F{\`a}brega}, Santi and {Schirmer}, Mischa and {Grebel}, Eva K. and {Saifollahi}, Teymoor and {Kanipe}, Jeff and {G{\'o}mez-Flechoso}, M. Angeles and {Akhlaghi}, Mohammad and {Javanmardi}, Behnam and {Wu}, Gang and {Eskandarlou}, Sepideh and {Bomans}, Dominik J. and {Henkel}, Cristian and {Block}, Adam and {Hanson}, Mark and {Schedler}, Johannes and {Teuwen}, Karel and {GaBany}, R. Jay and {Iba{\~n}ez Perez}, Alvaro and {Crawford}, Ken and {Promper}, Wolfgang and {Jimenez}, Manuel and {Farr{\`a}s-Aloy}, S{\'\i}lvia and {Mir{\'o}-Carretero}, Juan},
  title         = {{Stellar tidal streams around nearby spiral galaxies with deep imaging from amateur telescopes}},
  journal       = {\aap},
  year          = {2025},
  volume        = {701},
  pages         = {A182},
  month         = sep,
  adsurl        = {https://ui.adsabs.harvard.edu/abs/2025A&A...701A.182M},
  archiveprefix = {arXiv},
  doi           = {10.1051/0004-6361/202554980},
  eid           = {A182},
  eprint        = {2504.02071},
  primaryclass  = {astro-ph.GA},
}

@Article{Miguel13,
  author        = {{Cervi{\~n}o}, Miguel},
  title         = {{The stochastic nature of stellar population modelling}},
  journal       = {\nar},
  year          = {2013},
  volume        = {57},
  number        = {5},
  pages         = {123-139},
  month         = nov,
  adsurl        = {https://ui.adsabs.harvard.edu/abs/2013NewAR..57..123C},
  archiveprefix = {arXiv},
  doi           = {10.1016/j.newar.2013.09.001},
  eprint        = {1312.0015},
  primaryclass  = {astro-ph.IM},
}

@Article{McGaugh14,
  author        = {{McGaugh}, Stacy S. and {Schombert}, James M.},
  title         = {{Color-Mass-to-light-ratio Relations for Disk Galaxies}},
  journal       = {\aj},
  year          = {2014},
  volume        = {148},
  number        = {5},
  pages         = {77},
  month         = nov,
  adsurl        = {https://ui.adsabs.harvard.edu/abs/2014AJ....148...77M},
  archiveprefix = {arXiv},
  doi           = {10.1088/0004-6256/148/5/77},
  eid           = {77},
  eprint        = {1407.1839},
  primaryclass  = {astro-ph.GA},
}

@Article{Bell01,
  author        = {{Bell}, Eric F. and {de Jong}, Roelof S.},
  title         = {{Stellar Mass-to-Light Ratios and the Tully-Fisher Relation}},
  journal       = {\apj},
  year          = {2001},
  volume        = {550},
  number        = {1},
  pages         = {212-229},
  month         = mar,
  adsurl        = {https://ui.adsabs.harvard.edu/abs/2001ApJ...550..212B},
  archiveprefix = {arXiv},
  doi           = {10.1086/319728},
  eprint        = {astro-ph/0011493},
  primaryclass  = {astro-ph},
}

@Article{Bell03,
  author        = {{Bell}, Eric F. and {McIntosh}, Daniel H. and {Katz}, Neal and {Weinberg}, Martin D.},
  title         = {{The Optical and Near-Infrared Properties of Galaxies. I. Luminosity and Stellar Mass Functions}},
  journal       = {\apjs},
  year          = {2003},
  volume        = {149},
  number        = {2},
  pages         = {289-312},
  month         = dec,
  adsurl        = {https://ui.adsabs.harvard.edu/abs/2003ApJS..149..289B},
  archiveprefix = {arXiv},
  doi           = {10.1086/378847},
  eprint        = {astro-ph/0302543},
  primaryclass  = {astro-ph},
}

@Article{Zibetti09,
  author        = {{Zibetti}, Stefano and {Charlot}, St{\'e}phane and {Rix}, Hans-Walter},
  title         = {{Resolved stellar mass maps of galaxies - I. Method and implications for global mass estimates}},
  journal       = {\mnras},
  year          = {2009},
  volume        = {400},
  number        = {3},
  pages         = {1181-1198},
  month         = dec,
  adsurl        = {https://ui.adsabs.harvard.edu/abs/2009MNRAS.400.1181Z},
  archiveprefix = {arXiv},
  doi           = {10.1111/j.1365-2966.2009.15528.x},
  eprint        = {0904.4252},
  primaryclass  = {astro-ph.CO},
}

\clearpage
\begin{appendix}

\section{Young semi-resolved populations}\label{app:young}

\begin{strip}
\centering
\includegraphics[width=0.95\textwidth]{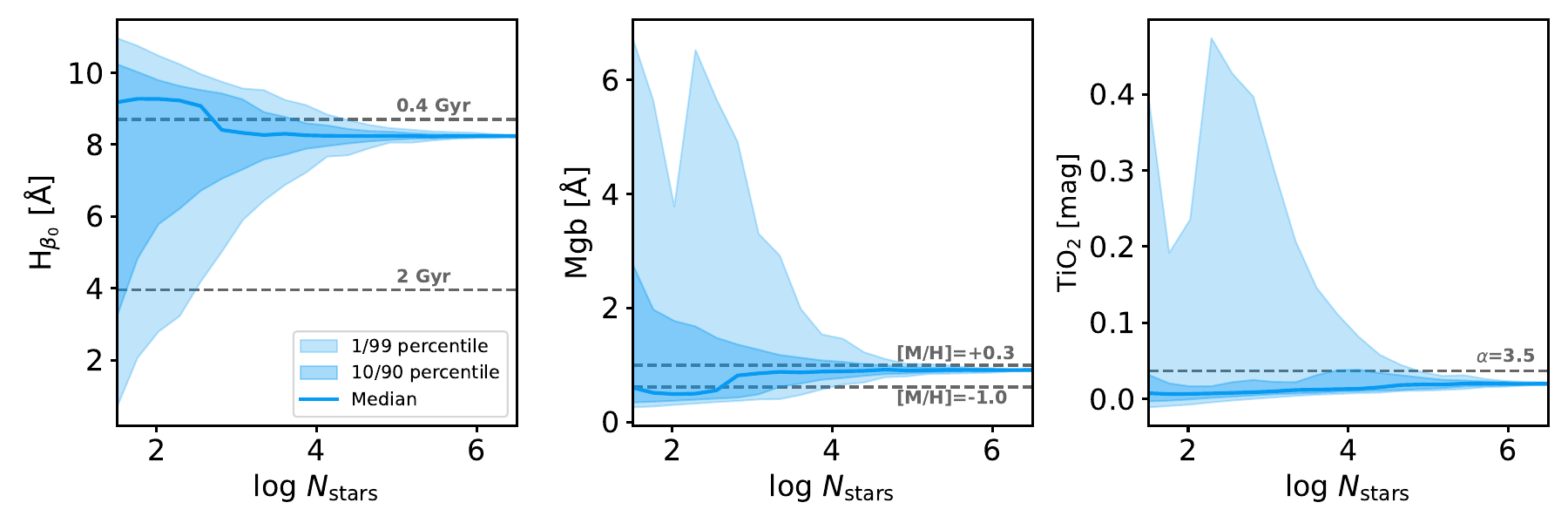}

\refstepcounter{figure}
\label{fig:young_indices}
{\small Fig.~\thefigure. Same as Fig.~\ref{fig:indices} but the line-strength predictions have been calculated assuming a 0.25 Gyr old population.}
\end{strip}

Figures~\ref{fig:young_indices} and ~\ref{fig:young_color} replicate Figs.~\ref{fig:indices} and~\ref{fig:color} from the main text but in this case for a young (0.25 Gyr) population, with solar metallicity and Milky Way-like IMF. Shaded contours indicate again the 1/99 and 10/90 percentiles of the distribution with a median value indicated by the solid line. Horizontal dashed lines in both cases mark the expected value for fully sampled SSP models with different ages, metallicities and/or IMF slopes.

The general behavior is similar as for old populations, with an evident increase in the scatter of the FASTAR predictions when the number of stars falls below $\log N_\mathrm{stars}\lesssim5$. However, the systematic biases in the colors and line-strengths are not the same as, for example, H$_{\beta_O}$ tends to increase with decreasing number of stars. Moreover, the apparent lack of a systematic bias in these indices for a Milky Way-like IMF seeing for old populations does not hold true for younger ages.

 \begin{figure}[!b]
    \centering
    \includegraphics[width=8cm]{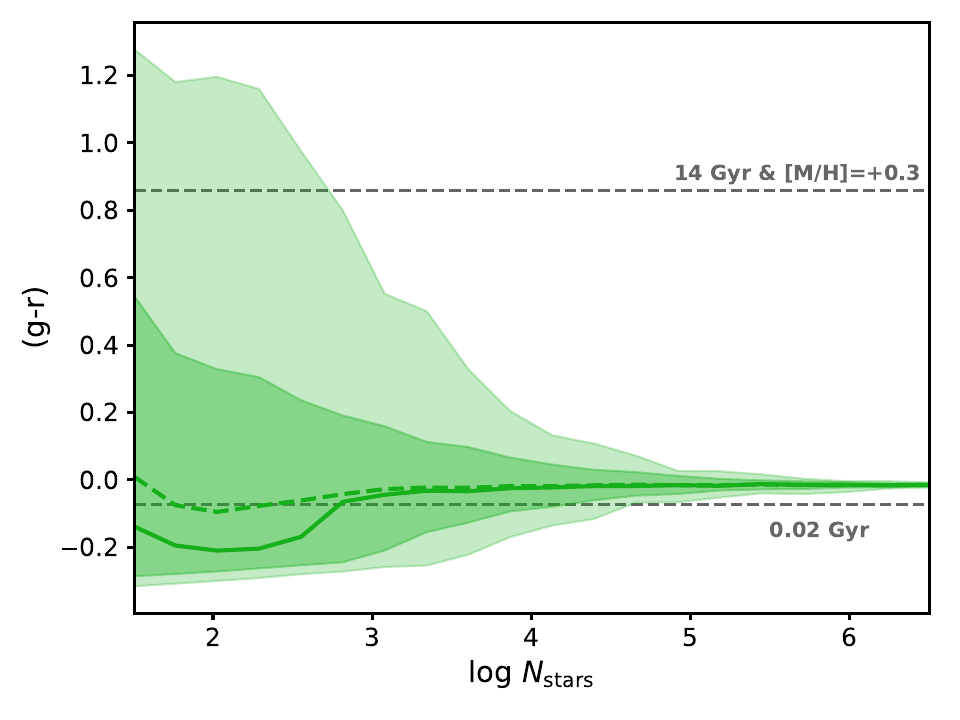}
    \caption{Same as Fig.~\ref{fig:young_color} but the $(g-r)$ color in this case corresponds to a younger (0.25 Gyr) stellar population.}
    \label{fig:young_color}
 \end{figure}

In order to further exemplify the behavior of semi-resolved FASTAR predictions for younger ages, Fig.~\ref{fig:hbeta_trend} shows the age dependence of the H$_{\beta_O}$ line for the three different $\log N_\mathrm{stars}$ ranges ($\log N_\mathrm{stars}=2.5$, 3.5 and 5.5, the same as in Fig.~\ref{fig:index_grid}).These model predictions assume a Milky Way-like IMF. For clarity, we only include the distribution of the 1/99 percentiles, indicated with colored shaded areas, with the median value corresponding to the solid lines.

In general, the behavior is similar for the three $\log N_\mathrm{stars}$ values, although there are clear systematics. For younger ages, low $\log N_\mathrm{stars}$ values tend to correspond to stronger H$_{\beta_O}$ values, while for older ages the trend is reversed. The bottom panel of  Fig.~\ref{fig:hbeta_trend} highlight these systematic differences showing the residuals with respect to the fully sampled FASTAR predictions. It is worth emphasizing that these systematic differences result from the biased sampling of specific stellar evolutionary phases, i.e., different stellar masses, and thus they are modulated by the assumed IMF.

 \begin{figure}[!b]
    \centering
    \includegraphics[width=8cm]{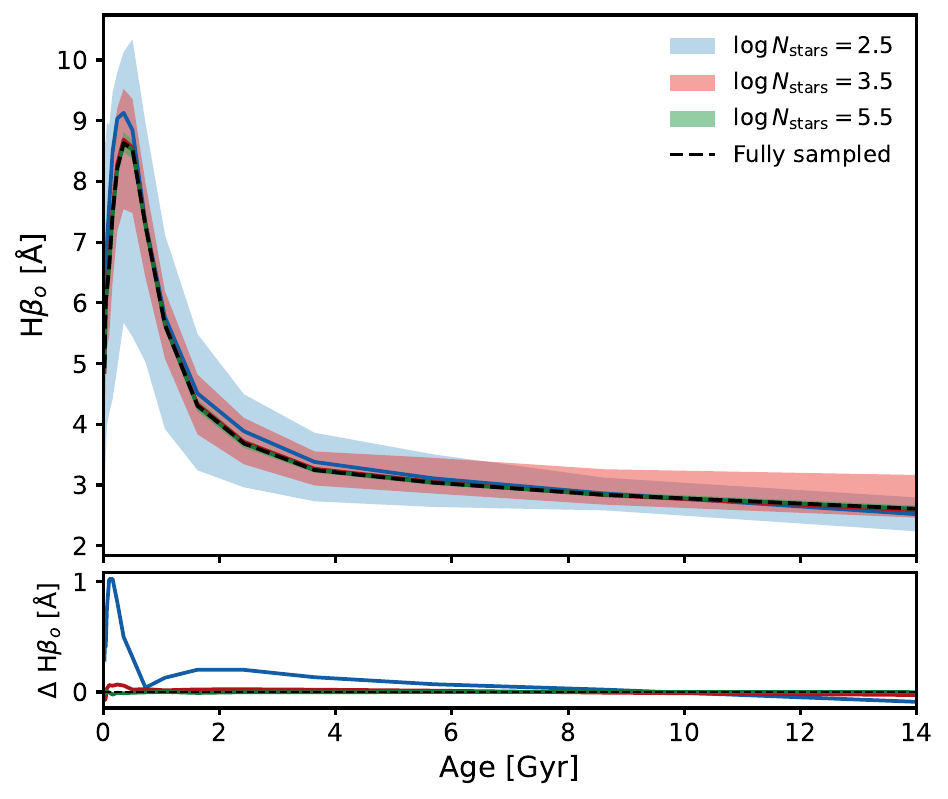}
    \caption{Age dependence of H$_{\beta_O}$. The top panel shows the distribution FASTAR H$_{\beta_O}$ predictions as a function of age, for three different $\log N_\mathrm{stars}$ values: 2.5 in blue, 3.5 in red and 5.5 in green. The shaded areas correspond to the 1/99 percentiles and the solid colored lines to the median trend. For reference, the predicted trend for a fully sampled population is shown with a dashed black line. The bottom panel shows the residuals between this dashed black lines and each of the median trends.}
    \label{fig:hbeta_trend}
 \end{figure}

\end{appendix}

\end{document}